%% file: Keen0512.tex
\documentclass{emulateapj}

\usepackage{graphicx} \usepackage{natbib} 
\shorttitle{The NIR LF Normalization at z~$\sim 0.1-0.3$} \shortauthors{Keenan et al.}

\begin{document}

\title{Testing for a large local void by investigating the Near-Infrared Galaxy Luminosity Function}
\author{R. C. Keenan\altaffilmark{1},
A. J. Barger\altaffilmark{2,3,4}, L. L. Cowie\altaffilmark{4},
W.-H. Wang\altaffilmark{1}, I. Wold\altaffilmark{2}, L. Trouille\altaffilmark{5}}

\altaffiltext{1}{Institute of Astronomy and Astrophysics, Academia Sinica,
  P.O. Box 23-141, Taipei 10617, Taiwan.}
\altaffiltext{2}{Department of Astronomy, University of Wisconsin-Madison, 475
N. Charter Street, Madison, WI 53706} 
\altaffiltext{3}{Department of Physics
and Astronomy, University of Hawaii, 2505 Correa Road, Honolulu, HI 96822}
\altaffiltext{4}{Institute for Astronomy, University of Hawaii, 2680 Woodlawn
Drive, Honolulu, HI 96822} 
\altaffiltext{5}{Northwestern University, 2145 Sheridan Road, Evanston, IL 60208-3112}

\begin{abstract} 
Recent cosmological modeling efforts have shown that a local underdensity on scales of a few hundred Mpc (out to $z \sim 0.1$), could produce the apparent acceleration of the expansion of the universe observed via type Ia supernovae.  Several studies of galaxy counts in the near-infrared (NIR) have found that the local universe appears under-dense by $\sim 25-50\%$ compared with regions a few hundred Mpc distant.  Galaxy counts at low redshifts sample primarily $L \sim L^*$ galaxies.  Thus, if the local universe is under-dense, then the normalization of the NIR galaxy luminosity function (LF) at $z>0.1$ should be higher than that measured for $z<0.1$.  Here we present a highly complete ($> 90$\%) spectroscopic sample of $1436$ galaxies selected in the $H-$band ($1.6\mu$m) to study the normalization of the NIR LF at $0.1<z<0.3$ and address the question of whether or not we reside in a large local underdensity.  Our survey sample consists of all galaxies brighter than $18^{th}$ magnitude in the $H-$band drawn from six widely separated fields at high Galactic latitudes, which cover a total of $\sim 2$ deg$^2$ on the sky.   We find that for the combination of our six fields, the product $\phi^* L^*$ at $0.1 < z < 0.3$ is  $\sim 30\%$ higher than that measured at lower redshifts.  While our statistical errors in this measurement are on the $\sim 10\%$ level, we find the systematics due to cosmic variance may be larger still.  We investigate the effects of cosmic variance on our measurement  using the COSMOS cone mock catalogs from the Millennium simulation and recent empirical estimates of cosmic variance derived by \citet{Driv10}.  We find that our survey is subject to systematic uncertainties due to cosmic variance at the $15\%$ level ($1~\sigma$), representing an improvement by a factor of $\sim 2$ over previous studies in this redshift range.  We conclude that observations cannot yet rule out the possibility that the local universe is under-dense at $z<0.1$.  The fields studied in this work have a large amount of publicly available ancillary data and we make available the images and catalogs used here.  

\end{abstract} 
\maketitle

\section{Introduction}
\label{intro}
The size of large-scale structures in the local universe (e.g. sheets, voids, and superclusters of galaxies), and our location among them, is of critical importance to the interpretation of observational results.  Cosmic variance due to large-scale structure can lead to systematic variations in observational data.  Such systematics can dominate over other sources of error if a volume sufficient to average over cosmic variance has not been sampled.  However, it remains unclear just what the upper limit on the size of large-scale structure is, and hence what volume constitutes a representitive sample of the universe.   If the typical size of local large-scale structure is much greater than $100~$Mpc, then local measurements of the Hubble constant, and other cosmological observables, could still harbor large systematic errors.

Cold dark matter simulations that include a cosmological constant ($\Lambda$CDM models), such as the Millennium Run \citep{Spri05}, predict
that the largest dark matter structures in the universe should be $\sim 100$ Mpc in extent.  Observed luminous matter large-scale structures, most notably the $>400~$Mpc Sloan Great Wall \citep{Gott05}, demonstrate the existence of structure on larger scales than simulations predict (assuming luminous matter traces dark matter).  It has also been shown that voids on scales similar to that of the Sloan Great Wall may explain the ``cold spots'' in the cosmic microwave background \citep{Inou06}.  It is not yet clear whether such structures represent extreme non-linearities in the matter distribution of the universe, or if inhomogeneities on several hundred Mpc scales are typical.  

Cosmologists have proposed so called ``void models'' as alternatives to $\Lambda$CDM, that invoke a large local underdensity to explain the apparent acceleration of the expansion of the universe \citep{Alne06, Chun06, Enqv07, Garc08a,Alex09,Garc09,Febr10,Cele10, Bisw10,Marr10,Clar10,Bole11a}.  The scale of the voids proposed in these models ranges from a few hundred Mpc to several Gpc in radius.  The basic premise of void models is that if we, as observers, live near the center of a large underdensity, then we would witness a local expansion of the universe that is faster than the global expansion, simply due to the void being evacuated toward higher-density surroundings via gravity.  This would provide for a locally measured Hubble constant that is higher than the global value and look observationally like an accelerating expansion.  

 In their current form, void models with $\Lambda = 0$ appear not to be viable alternatives to $\Lambda$CDM, as they have trouble fitting the entire range of cosmological observables \citep{Garc08b, Zibi08, Moss11, Zhan11, Ries11}.  However, the exploration of this class of cosmological models, and other inhomogeneous models, has highlighted the need for a more comprehensive understanding of large-scale structure in the universe and our location within it (e.g. \citealt{Marr11,Bole11b,Bull12}).  In particular, so called ``minimal void'' scenarios (e.g., \citealt{Alex09, Bole11a}) have shown that very simple models that place the observer near the center of a void that is $\sim 250~h^{-1}$~Mpc in radius (to $z \sim 0.1$) and $\sim 50\%$ under-dense compared to its surroundings are sufficient to explain the apparent acceleration observed via type Ia supernovae.  While these models are simplistic, they point out that our location within local structure may have profound implications for our measurement of cosmological observables.  

The existence of a large local void would be consistent with studies of near-infrared (NIR) galaxy counts that have found the local space density of galaxies may be low by $25-50$\% compared to the density at distances of  $\sim 300~h^{-1}$~Mpc or $z \sim 0.1$ \citep{Keen10a,Buss04,Frit03,Frit05,Huan97}.   Galaxy counts in the NIR at low redshifts are primarily sampling $L \sim L^*$ or brighter galaxies \citep{Barr09}.  Thus, if the universe at $z<0.1$ is under-dense, the normalization of the NIR luminosity function (LF) at $z>0.1$ should be higher than locally measured values.   In this paper we explore the normalization of the NIR LF as a function of redshift to test for the existence of a large local underdensity.

At $z<0.1$, the NIR LF has been studied using samples selected from large NIR surveys, such as the Two Micron All Sky Survey (2MASS, \citealt{Skru06}) and the UKIRT Infrared Deep Sky Survey (UKIDSS, \citealt{Lawr07}), combined with spectroscopy from redshift surveys, such as the CfA2 redshift survey \citep{Gell89,Huch92}, the Two-degree Field Galaxy Redshift Survey (2dFGRS, \citealt{Coll01}) and the Sloan Digital Sky Survey (SDSS, \citealt{York00}). 

Using these data, the $\langle z \rangle \sim 0.05$ NIR LF has been relatively well established \citep{Cole01, Koch01, Jone06}.  Studies at slightly higher median redshifts ($0.07<z<0.1$) tend to arrive at a normalization that is a factor of $\sim 1.5$ higher, though they state general consistency with lower redshift measurements given possible systematics due to differences in methodology, etc. \citep{Bell03, Eke05, Smit09, Hill10}.  Taken together, however, the aforementioned studies appear to point to the possibility of an increasing LF normalization from $z\sim 0.05$~to~$z\sim 0.1$.  

Ideally, one would like to study the total luminosity density (and hence stellar mass density) as a function of distance, rather than just simply the normalization of the LF.  However, any apparent magnitude limited survey at low redshifts is limited on the bright end of the LF by poor counting statistics and on the faint end by the relatively small(er) volume over which faint galaxies may be sampled.  These two effects, along with differences in methodology, combine to yield the result that the overall shape of the NIR LF can be quite different from one study to the next.  These differences in LF shape can lead to substantial differences in integrated luminosity density.  
 
 The peak contribution of the LF to the total luminosity density occurs at $L \sim L^*$, where all studies feature the best statistics.  The Schechter (1976) function paramters for the normalization ($\phi^*$) and characteristic luminosity ($L^*$) are correlated.  Thus, in this study we focus on a comparison between our study and those from the literature, of the product of the normalization and characteristic luminosity ($\phi^* L^*$), which amounts to a comparison of the peak of the luminosity density distribution across studies.  

In this paper, we probe the NIR LF just beyond the local volume at redshifts of $0.1<z<0.3$.  To study the NIR LF at these relatively low redshifts, high spectroscopic completeness is essential to overcome possible biases and to deal with the fact that the errors in the photometric redshifts are of the order of the redshifts themselves (i.e. $\sigma_z / z \sim 1$).  Our spectroscopic survey of $H-$band selected galaxies ($H_{\rm{AB}}<18$) is $>90\%$ complete over six widely separated fields covering a total of $2$ deg$^2$ on the sky.  Rest-frame $H-$band light is a good tracer of stellar mass and hence a galaxy's $H-$band luminosity (in solar units) is approximately equal to its stellar mass (in solar units)~ \citep{Dejo96,Bell01,Bell03,Kirb08}.  At low redshifts, $\lambda_{observed} \approx \lambda_{rest}$, and, in the NIR, $K-$corrections are small and nearly independent of galaxy type \citep{Mann01}, so this is effectively a mass-selected sample.

Our survey is unique among extragalactic redshift surveys because of its relative depth, width, and spectroscopic completeness.   Deep redshift surveys, such as the Cosmic Evolution Survey (COSMOS / ZCOSMOS, \citealt{Scov07a,Lill09}), the All-wavelength Extended Groth strip International Survey plus the Deep Evolutionary Exploratory Survey 2 (AEGIS + DEEP2, \citealt{Davi07}), and the  VIMOS VLT Deep Survey (VVDS, \citealt{Lefe05}) all feature high-quality NIR photometry  and $\sim 10,000$ publicly available redshifts each.  However, the spectroscopic completeness at relatively bright NIR magnitudes ($H_{\rm{AB}}<18$) is less than $40\%$ in all cases.   The Galaxy And Mass Assembly survey (GAMA, \citealt{Driv11}) will feature high completeness to $K\sim18$, although the current public data release is only complete to $K\sim 16$.

The structure of this paper is as follows:  We present our observations, data reduction, and redshift determination methods in Section~\ref{obsred}.  We discuss our methods of spectral energy distribution (SED) fitting to obtain photometric redshifts for targets lacking spectroscopy in Section~\ref{GAZELLE}.  We present our results regarding the observed NIR LFs in Section~\ref{lumfunc}.  We analyze the effects of cosmic variance on our study and the last decade of studies of the NIR LF in the literature in Section~\ref{cv}.  We summarize in Section~\ref{summary}.  Unless otherwise noted, all magnitudes given in this paper are in the AB magnitude system ($m_{\rm{AB}} = 23.9-2.5~\rm{log}_{10}~$$ f_\nu$ with $f_\nu$ in units of $\mu$Jy).  We assume a cosmology of $\Omega_M = 0.27, \Omega_{\Lambda} = 0.73,~$and$~h = 0.7$ in our conversion of redshifts to distances.

\section{Observations and Data Reduction}
\label{obsred}
\input{Table1}

Our primary photometry covers six fields in the $J,H,$~and~$K-$bands. In Table 1 we show the location of our observed fields in right ascension and declination, as well as Galactic and supergalactic coordinates.  Also shown in Table 1 is the area on the sky, number of target galaxies, and spectroscopic completeness for each field.

We made all the $H$ and $J-$band observations (except the CDF-N $J-$band) with the Ultra Low Background Camera (ULBCam) on the UH 2.2~m telescope.  We observed the CDF-N in the $K_s-$band with the Widefield Infrared Camera (WIRCam) at the Canada France Hawaii Telescope
3.6~m (CFHT).  CDF-N $J-$band observations with WIRCam were obtained by a group led by Lihwai Lin in 2006A.  Our group reduced these public data in 2008.  We observed five of our six fields in the $K-$band with WFCam on UKIRT 

We presented the observations, data reduction, star-galaxy separation, and
bright galaxy counts for our deep wide-field NIR imaging campaign in \citet{Keen10a}.  In that paper we used the bright NIR galaxy counts from our survey, in combination with other data from the literature, to explore local large-scale structure via the slope of the galaxy counts curve as a
function of position on the sky.  In \citet{Keen10b} we integrated galaxy counts from our NIR survey in combination with deeper data from the Multi Object InfraRed Camera and Spectrograph (MOIRCS) instrument on the Subaru Telescope, and other studies from the literature, to obtain the best current estimate of the total light from galaxies in the NIR.   

The portion of the survey used for this work includes only areas where all three bands have uniform coverage and consists of
approximately 2 deg$^2$ reaching a 5~$\sigma$ limiting magnitude of $JHK
\sim 22-23$ over $\sim 1.8$ deg$^2$ with another $\sim 0.2$ deg$^2$ to
$JHK \sim 24$.  Thus, the NIR photometry used for the present work has very high signal to noise. 

Before performing photometry on galaxies, we calibrated our NIR images to 2MASS fluxes using relatively bright  point sources in the magnitude range $14 < JHK < 16$, where 2MASS reports better than $10~\sigma$ signal to noise and our data are well below saturation levels.  We found our final calibrated images to be quite flat and free from systematic variations in flux as a function of position.  We used the SExtractor \citep{BA96} MAG\_AUTO aperture to do photometry for galaxies in this study.  We found that this aperture did an excellent job of extracting total magnitudes for high signal to noise galaxies (such as those used in this study).  For a detailed description of the data reduction and calibration procedures used, we refer the reader to \citet{Keen10a}.  

Here we selected all galaxies brighter than 18$^{th}$ magnitude in the $H-$band in our fields and targeted them in a campaign of spectroscopic follow-up.  As noted in Section~\ref{intro}, this is essentially a mass-selected sample.  We note that thermally pulsating asymptotic giant branch (TP-AGB) stars may dominate the NIR light in young ($0.2-2~$Gyr) stellar populations \citep{Mara06}, which may be a source of error in this selection for galaxies that contain a significant population of young stars.  However, \citet{Zibb12} have recently found that post-starburst galaxies at $z \sim 0.2$ show NIR fluxes, relative to optical, that are consistent with \citet{Bruz03} stellar population models, which suggests TP-AGB stars may not have as significant of an impact as previously believed.

\subsection{Fields Observed}
\label{fields}
Our first two fields are centered on the \emph{Chandra} Large Area
Synoptic X-ray Survey (CLASXS; \citealt{Yang04,Stef04}) and the \emph{Chandra} Lockman Area North Survey
(CLANS; \citealt{Trou08,Trou09}). Each of these fields cover $\sim 0.5$~deg$^2$ in
$JHK$.  These fields  are located in the Lockman Hole region of
low Galactic HI column density \citep{Lock86}.  Our third field covers a
$\sim 0.2~$deg$^2$ area centered on the \emph{Chandra} Deep Field North (CDF-N, \citealt{Bran01, Alex03}).  The CDF-N contains the Great Observatories Origins
Deep Survey North (GOODS-N; 145 arcmin$^2$ \emph{HST}
Advanced Camera for Surveys observation, \citealt{Giav04}).  Our
fourth field is the Abell 370 (A370) cluster and surrounding area ($\sim 0.4$
deg$^2$).  A370 is a cluster of richness 0 at a redshift of $z=0.37$. Our
fifth and sixth fields are $\sim 0.2$ deg$^2$ each centered on the ``Small-Survey-Area 13''(SSA13) and  ``Small-Survey-Area 17'' (SSA17) from the Hawaii Deep Fields described in \citet{Lill91}. 

\subsection{Ancillary Photometry}
All the fields included in this work (except A370) are covered by the SDSS, which allows us to use uniform photometry for fitting SEDs.  In addition to the fields that were presented in \citet{Keen10a} (CLANS, CLASXS, CDF-N, A370, and SSA13), we added the blank field ``SSA17''.  The NIR data reduction and
photometry methods for the SSA17 field were identical to those described in
\citet{Keen10a} for the other fields, except SSA17 does not include $K-$band photometry.  

In the case of the A370 field, we use publicly available optical
photometry from the Canada France Hawaii Telescope (CFHT) MegaCam in the
$u,g,r,$~and~$i$ bands (there were no $z-$band observations for A370).  We downloaded the catalogs from the MegaCam image stacking pipeline website (Megapipe\footnote{http://cadcwww.dao.nrc.ca/megapipe}).
The MegaCam $u,g,r,~$and$~i~$filters are quite similar to those of the
SDSS\footnote{http://cadcwww.dao.nrc.ca/megapipe/docs/filters.html}, and small
magnitude offsets provided on the Megapipe website bring the data into accordance with SDSS magnitudes.  

Catalogs generated through Megapipe are processed in the following way:  Raw images are reduced or ``detrended'' through the Elixir\footnote{http://www.cfht.hawaii.edu/Instruments/Elixir} system, which includes treatment for dark and bias subtraction, flatfielding, defringing, and basic astrometric and photometric calibration.  Elixir processed images are retrieved and checked for quality before stacking.  More refined processing for astrometric and photometric calibrations are performed and then the images are coadded.  Catalogs are generated through the application of the source extraction software  SExtractor \citep{BA96}.  For a detailed description of all processes in the Megapipe reduction, please refer to the website (http://cadcwww.dao.nrc.ca/megapipe). 

Three of our fields (CLANS, CLASXS, and CDF-N) are covered in part by
\emph{Spitzer} InfraRed Array Camera (IRAC $3.6 - 8~\mu$m,~\citealt{Fazi04})
campaigns.  In the CLANS and CLASXS fields, we use the Spitzer Wide-area
InfraRed Extragalactic Survey catalogs (SWIRE,~\citealt{Lons03}) to supplement
the photometry for 516 of our target galaxies.  In the CDF-N, we use the IRAC
catalogs of \citet{Wang10} to supplement the photometry for an additional 33 targets.  As a result, almost $40\%$ of our target galaxies have high quality photometry in the mid-infrared.  

\subsection{Spectroscopic Observations and Reductions}

We observed our target galaxies between 2006 and 2010 with the Hydra fiber
spectrograph instrument \citep{Bard95} on the Wisconsin-Indiana-Yale-NOAO
(WIYN) telescope at Kitt Peak in Arizona.  Hydra is a multi-fiber spectrograph
with $\sim 90$ functioning fibers in the configuration we used.  The field of view of this instrument is 1
degree in diameter on the sky and fibers may be placed a minimum of
$37\arcsec$ apart.  We configured the fibers using the \emph{whydra}
configuration code, which allows the user to define a minimum number of sky positions
and guide stars required and allows for targets to be weighted by priority.
We also used the \emph{hydrasim} simulator code to check that the output of
\emph{whydra} was compatible with the software at the telescope.  The hydra
``red'' fibers are $2\arcsec$ in diameter, which was ideally suited to getting
the majority of light from our target galaxies onto the spectrograph with a
minimum of contamination from the sky and neighboring sources.  

For this survey we selected galaxies from our $H-$band photometry
($H<18$).  We masked areas of our images near bright objects ($R_{magnitude} < 14.5$), as
described in \citet{Keen10a}, to avoid targets that were blended with bright neighbors.
This selection yielded $1436$ target galaxies in the range $14.5 < H <
18$. 

We configured the spectrograph using the ``red'' fiber bundle
and the 316@7.0 grating at first order with the GG-420 filter to provide a
spectral window of $\sim 4000 - 9500$~\AA~with a pixel scale of $2.6$~\AA~per
pixel.  Toward the edges of our spectral window the spectra were often
dominated by noise due to the waning sensitivity of the spectrograph on the
blue end and the spectrum becoming sky-dominated in the red.  As a result, the
useable spectral window was $\sim 4500 - 9000$~\AA. We obtained the majority of our target spectra since 2008 when the bench spectrograph at WIYN was upgraded \citep{Bers08}.   The improved sensitivity of the upgraded bench spectrograph allowed us to determine redshifts for $\sim~90-100\%$ of target galaxies in a two hour exposure given ideal observing conditions.

Our exposures were typically 20 minutes each, and we would stay on target for
1-3 hours total, depending on the observing conditions and the targets observed.  For calibration we took dark frames, bias (zero exposure time) frames, dome flats, and CuAr
comparison lamp exposures.  We employed the iraf task \emph{dohydra}\footnote{http://iraf.noao.edu/tutorials/dohydra/dohydra.html} in the reduction of our
spectra.  This task is specifically designed for reduction of data from the
Hydra spectrograph and includes steps for dark and bias subtraction,
flatfielding, dispersion calibration, and sky subtraction.

\subsection{Redshift Determination}

\label{redshift}

\begin{figure}
\begin{center}
\includegraphics[width=90mm]{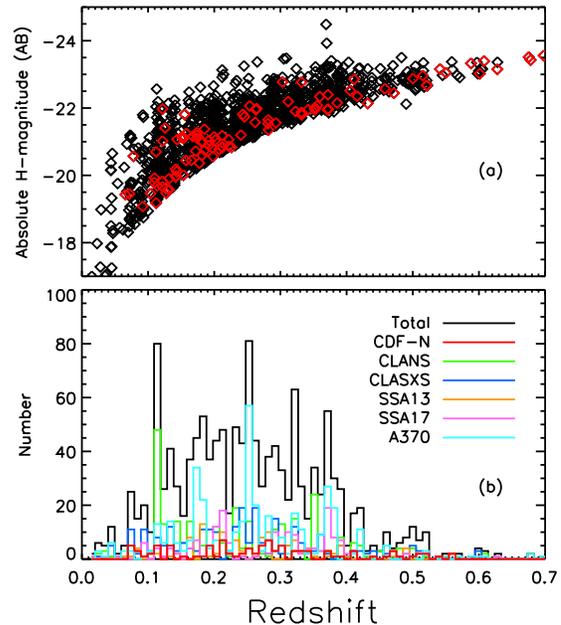}
\caption{\label{zhist}  (a) $H-$band absolute magnitude versus redshift for all galaxies in our sample.  Black points represent the spectroscopic redshift sample and red points represent the photometric redshift sample. (b) All redshifts in our six fields.  The total for all six fields is shown as a black histogram and each field is represented individually with colored histograms.}
\end{center}
\end{figure} 

We used the iraf task \emph{xcsao} within the \emph{rvsao} package
\citep{Kurt98} to determine redshifts for our observed spectra.  The
\emph{xcsao} task is an interactive tool that allows the user to determine
redshifts by cross-correlating a spectrum against template spectra using
methods described in \citet{Tonr79}.  This task allows the user to manually
mask portions of the spectrum and cycle through a variety of candidate fits to
find the best match.  We manually determined a redshift for all of the
galaxies in our spectroscopic sample using \emph{xcsao} in this manner.  

At the relatively low redshifts of our sample galaxies, the commonly identifiable absorption features in our spectra were CaII H~$(3969$~\AA) and K~$(3934$~\AA), the 4000~\AA~break, the Fe G-band feature $(\sim 4100$~\AA), H$\gamma~(4342~$\AA), H$\beta~(4865~$\AA), and the Na D lines $(5890/5896~$\AA).  If emission lines were present, our spectra typically captured the H$\alpha~(6563~$\AA) and [NII] $(6548/6584~$\AA) complex along with
the [SII] doublet $(6719/6730~$\AA)~and/or the [OII] $(3727~$\AA), [OIII] $(4961/5007~$\AA) and
H$\beta~(4865~$\AA) emission lines.

For $\sim 50$ galaxies we targeted objects that already had a cataloged
redshift in the SDSS archive.  We find 100\% agreement with the SDSS redshifts
to within an rms error of 0.0005.  We supplement our catalogs with an
additional 32 redshifts from the SDSS for other targets in our
fields. We also supplement our catalogs with 129 redshifts for
objects in our fields that were either previously published in the literature
or obtained from other spectroscopic campaigns within our research group.  

For the CDF-N field, many of these supplementary redshifts are cataloged in \citet{Barg08} and references therein.  In the CLANS and CLASXS fields, many of the supplementary redshifts are cataloged in \citet{Trou08, Trou09}.  Many redshifts in the SSA13 field are cataloged in \citet{Cowi04} and in the A370 field in I. Wold et al. (2012, in preparation).  The remainder come from an ongoing campaign of followup spectroscopy in our fields using the DEep Imaging Multi-Object Spectrograph (DEIMOS) on the Keck telescope (L. Cowie, private communication).  

In Figure~\ref{zhist}a, we show $H-$band absolute magnitude versus redshift, where black points represent the spectroscopic sample and red points represent photometric redshifts (see Section~\ref{GAZELLE}). In Figure~\ref{zhist}b, we show the spectroscopic redshifts in our fields.  The black histogram shows the total for all six fields, and the colored histograms show each field individually. 

For our redshifts obtained with the Hydra instrument (1155 total), we assigned a simple confidence flag to the determined redshift to indicate whether it was a high-confidence redshift (1) or questionable (0).  In reality there may be some ambiguity, in relatively low signal to noise spectra, as to when one can robustly determine a redshift versus when a redshift may be determined but could be questionable.  To keep things simple in this case, we assign high confidence to redshifts where a minimum of two absorption or emission features could be clearly identified in different parts of the spectrum simultaneously.  We assign questionable confidence when one or more features appear identifiable in a spectrum, such that a redshift could be defined, but that other possible fits to the spectrum cannot be completely ruled out.  With this system we assign high-confidence to $\sim 92\%$ of the redshifts in our Hydra sample.

\section{Spectral Energy Distribution Fitting Using \emph{GAZELLE}}
\label{GAZELLE}

We used the \emph{GAZELLE} code (R. Kotulla et al. 2012, in preparation) to fit the SEDs of our target galaxies in order to obtain photometric redshifts for galaxies that lack spectroscopy.  The \emph{GAZELLE} code uses the GALaxy EVolution (\emph{GALEV}, \citealt{Kotu09}) stellar population synthesis models as the basis for fitting SEDs. \emph{GALEV} is an evolutionary synthesis code that allows the user to generate models to describe the evolution of stellar populations in general, whether they be resolved populations (as in the case of star clusters) or unresolved populations in galaxies.  

Given a set of initial conditions, \emph{GALEV} is able to track the integrated light properties of galaxies over cosmological timescales. Users are now able to generate their own models directly via the \emph{GALEV} website (http://www.galev.org).  To generate a model, the user first chooses a galaxy type (E, Sa, Sb, Sc, Sd) or specifies a particular star formation history (e.g., constant, exponentially declining etc.), including an option to upload a user defined star formation history.

 \emph{GALEV} is unique among evolutionary synthesis codes in that it allows the user to select a ``chemically consistent'' treatment by simultaneously tracking the spectral evolution of stellar populations and the chemical evolution (metallicity) of the gas component of galaxies.  Alternatively, one can choose from five fixed metallicites in the range $-1.7 < \rm{[Fe/H]} < +0.3$.  The user may also vary the initial mass function (IMF), switch on or off the emission from gas, choose a Galactic extinction prescription, and vary parameters regarding the details of star formation efficiency and burst duration, age, e-folding time, etc., for starburst models.  Lastly, the user chooses a cosmological parameter set, desired outputs of the model, and the photometric filter profiles to be applied to model SEDs.  

In an effort to fit model SEDs to photometric data, a wide selection of models is required to cover a range of star formation histories, ages, metallicities, galaxy types, and so on.  In this work, we chose 83 independent \emph{GALEV} models to use in fitting our photometry.  20 of these models are the types available through the \emph{GALEV} website for five galaxy types (E, Sa, Sb, Sc, and Sd).  Each galaxy type is modeled at three fixed metallicities ($\rm{[Fe/H]} = -0.7, -0.3, 0$) and once in the ``chemically consistent'' evolving metallicity format.  The other 63 models are starburst models that attempt to capture a wide range in post-burst age.  The starburst models are not publicly available through the \emph{GALEV} website.  They are described in detail in the dissertation of Ralf Kotulla and were provided to us directly by him.  Altogether, we use burst models covering 21 age steps (post-burst 0.002-9 Gyr) at three fixed metallicities ($\rm{[Fe/H]} = -0.7, -0.3, 0$). 

\emph{GALEV} does not yet handle emission from dust, so that in low-redshift samples such as this work, the longer wavelength \emph{Spitzer} bands contribute relatively little to the SED fits.  This is because the \emph{GAZELLE} code ignores photometry in bands where dust emission is expected to dominate the signal ($> 3~\mu$m).  

\emph{GAZELLE} uses a $\chi^{2}$ algorithm to fit models to the observed data.  The $\chi^{2}$ values for each model are then transformed into normalized probabilities.  The model with the highest probability is the one chosen for computing redshift, mass, metallicity, and so on.  The formal $1~\sigma$~uncertainties in redshift, mass, metallicity etc. are derived by finding the minimum and maximum values associated with the models in the top $68\%$ of the normalized probability distribution.  

\emph{GAZELLE} includes an option to run with a user defined redshift for galaxies.  This mode allows the user to ``calibrate'' \emph{GAZELLE} given the difference between observed photometry and the output model photometry from the best-fit SED.  In this mode, we compared the output model magnitudes from \emph{GAZELLE} with the input observed magnitudes and found very good agreement in the optical ($|m_{observed}-m_{model}| < 0.03$), but systematic offsets of $\sim 0.1-0.2$ in the NIR.  We then included these offsets in the \emph{GAZELLE} input parameter file to account for these systematics, such that, when we re-ran the code, the model output magnitudes agreed with the input magnitudes in all bands with systematics of $<0.03$ magnitudes.  These adjustments served to improve the spectroscopic versus photometric redshift relation. 

At the relatively low redshifts of our sample, the error in photometric redshifts can be of the same order as the redshifts themselves.  Despite these relatively large uncertainties, we find that the photometric redshifts derived for galaxies that lack spectroscopy fall within the range of expected values from the spectroscopic sample.  We note that $> 50\%$ of our targets have a $|z_{spec}-z_{phot}| / (1+z) < 0.05$ and $>80\%$ have $|z_{spec}-z_{phot}| / (1+z) < 0.1$, which could be considered fairly typical for determining photometric redshifts.

\section{The NIR Luminosity Function}
\label{lumfunc}

The NIR LF of galaxies has been relatively well established at low redshifts of $\langle z \rangle \sim 0.025-0.06$ \citep{Cole01, Koch01, Jone06}.      At slightly higher redshifts of $\langle z \rangle \sim 0.06-0.1$, large surveys have also provided constraints on the LF and tend to arrive at a  normalization that is a factor of $\sim 1.5$ times higher than lower redshift studies  \citep{Bell03, Eke05, Smit09, Hill10}.   

At the mean redshift of our sample (z $\sim 0.2$) the LF is not particularly well constrained because, in this redshift range, depth and wide area are simultaneously required to sample the entire LF.  Furthermore, high spectroscopic completeness is essential at these redshifts to overcome possible biases, and to avoid the relative uncertainty in photometric redshifts ($\sigma_z / z \sim 1$), which lead to large uncertainties in absolute luminosities.

Here we combine our fields to study the NIR LF normalization at $0.1<z<0.3$ and to compare with lower redshift values.   As noted above, our sample is selected in the $H-$band, but we have photometry for the entire sample in the $J$ and $K-$bands as well (presented in \citealt{Keen10a}).  Thus, we are able to derive LFs in $J, H$~and~$K$ to compare with a wide range of selections from the literature.  

First, we derived LFs for the combination of all six of our fields in the $H-$band using four different estimators for both the LF and its normalization.  From these methods we devise a hybrid scheme of deriving the LF that combines two different LF estimators.  We compute the  $J, H,$~and~$K-$band LFs in this way to compare the overall shape and normalization with other studies from the literature.  In the range $0.1<z<0.3$, we have 812 galaxies. This subsample is $92\%$ spectroscopically complete.  The only other study to focus specifically on the NIR LF in this redshift range was performed by \citet{Feul03}.  Our sample represents a factor of four increase in the number of galaxies used in their work.  

\subsection{Methods for Calculating the Luminosity Function}

Several methods exist for calculating the LF of galaxies.  The most commonly assumed functional form of the LF is that of the \citet{Sche76} function, which takes the following form

\begin{equation}
\Phi (L)dL = \phi^*\bigg( \frac{L}{L_*}\bigg)^\alpha {\rm exp} \bigg( \frac{-L}{L_*} \bigg) \frac{dL}{L_*},
\end{equation}

\noindent which may be converted to be in terms of absolute magnitudes

\begin{equation}
\frac{L}{L_*} = 10^{-0.4(M-M^*)},
\end{equation}

\noindent such that

\begin{equation}
\Phi(M) = 0.4\rm{ln}(10)\phi^* \frac{\bigg(10^{0.4(M^*-M)}\bigg)^{(\alpha+1)}}{\rm{exp}({10^{0.4(M^*-M)})}}.
\end{equation}

The Schechter function parameter $L_*$ (or $M^*$) represents the luminosity of galaxies at the knee of the LF, while $\phi^*$ determines the number density of $L_*$ galaxies, and $\alpha$ is the faint-end slope.  Methods for deriving the LF include parametric  and non-parametric forms.  In this work we use a hybrid combination of parametric and non-parametric estimators to determine the Schechter function parameters, but first we compare four methods of LF estimation independently.  

\subsection{Parametric LF Estimation}
\label{parlf}
The most commonly used parametric method is that of \citet{Sand79} (STY).  This method employs a maximum likelihood estimator (MLE) and the assumption that the LF takes the Schechter function form.  An advantage of this method is that no binning is required.  Instead, a probability density, $p(M_i, z_i)$, is calculated for each galaxy with absolute magnitude $M_i$ at redshift $z_i$, where $M_{\rm{faint}}(z_i)$ is the faintest galaxy visible in an apparent magnitude limited survey at redshift $z_i$, such that

\begin{equation}
p(M_i, z_i) = \frac{\Phi (M_i) }{  \int_{-\infty}^{M_{\rm{faint}}(z_i)} \Phi (M^\prime ) dM^\prime}\,.
\end{equation}

\noindent The likelihood function is then the product of  the probability densities for all galaxies in the survey

\begin{equation}
\mathcal{L} = \prod_i p(M_i,z_i)\,.
\end{equation}

 This product is maximized to obtain the best fit Schechter function parameters.  In this method, the density functions cancel in the calculation of the probability densities.  Thus, this technique is insensitive to density inhomogeneities, and the normalization ($\phi^*$) must be determined via some other method.  
 
 \subsection{Non-parametric LF Estimation}
 \label{nonparlf}
 Non-parametric LF estimators include binned and non-binned methods, and no a priori assumption of the LF functional form is required.  However, for purposes of comparison with other studies and other methods a Schechter function is usually fit to the resulting LF estimation.
 
 The simplest non-parametric approach is the $1/V_{max}$ method of \citet{Schm68}.  This method assumes a homogeneous distribution of galaxies, such that each galaxy is assigned a weight ($1/V_{max}$) based on the fraction ($V_{max}$) of the total survey volume  in which that galaxy can be observed, given its absolute magnitude.  The sum of the weights for galaxies in an absolute magnitude bin is then the number density, $\Phi(M)$, for that bin.  This method has the advantage that the normalization of the LF is determined directly, but it has been shown that this method may underestimate the number density of galaxies near the flux limit of a sample \citep{Page2000}.  
 
 Another non-parametric approach is the $C^-$ method of \citet{Lynd71}.   The $C^-$ method does not require binning, and it does not require the assumption of a homogeneous distribution of objects.   Thus, as for STY, the normalization must be determined by other means. In the $C^-$ method, the cumulative luminosity function is given by
 
\begin{equation}
\phi(M_k) = \int_{-\infty}^M \Phi(M) dM = \psi_1 \prod_i^{M_k<M} \frac{C_k^-(M) +1}{C_k^-(M)}\,.
\end{equation}
 
 In a sample with a redshift lower limit of $z_{\rm{min}}$, the value for the parameter $C_k^-$ for each galaxy at $(M_k, z_k)$ is the total number of galaxies that are both brighter than $M_k$ in absolute magnitude and within the redshift range ($z_{\rm{min}} < z_k < z_{k,limit}$), where $z_{k,limit}$ is the maximum redshift that a galaxy with absolute magnitude $M_k$ may be observed in the survey.  In other words,
 
 \begin{equation}
 C_k^-(M_k) =  \sum_i^{N_{gal}} W(M_k)\,,
 \end{equation}
 
\noindent where the window function $W(M_k)$ is defined as
 
 \begin{equation}
 W(M_k) =  \left \{ \begin{array}{l} 1,~~~~M<M_k~~\rm{and}~~ z_{\rm{min}} < z_k < z_{k, limit} \\ 0,~~~~~\rm{otherwise} \end{array} \right.  \,.
 \end{equation} 
 
 The resulting cumulative luminosity function may then be binned and fit with a Schechter function for comparison with other studies.  
 
 Another popular non-parametric approach is the ``Step-Wise Maximum Likelihood'' estimator (SWML, \citealt{Efst88}).  This method requires binning but no a priori assumption of homogeneity.  In this case, the LF is parameterized into $N_p$ steps of width $\Delta M$ such that
 
 \begin{equation}
 \Phi(M)  = \Phi_k,~~~ k=1,...,N_p\,.
 \end{equation}
 
Two window functions, $W(x)$ and $H(x)$, are defined such that
 
 \begin{equation}
 W(x) =    \left \{ \begin{array}{l} 1,~~~~-\Delta M/2 \leq x \leq \Delta M/2 \\ 0,~~~~~\rm{otherwise} \end{array} \right. \,.
 \end{equation}
 
 \begin{equation}
H(x) =   \left \{ \begin{array}{ll} 1, & x > \Delta M/2 \\ (x/\Delta M + 1/2), & -\Delta M/2 \leq x \leq \Delta M/2 \\ 0, & x < -\Delta M/2 \end{array}  \right. \,.
 \end{equation}

Maximizing the likelihood function (please see \citealt{Efst88} for details) then yields the following form of the differential LF
  
 \begin{equation}
 \Phi_k(M) \Delta M = \frac{\sum_i^{} W(M_k-M_i)}{\sum_i^{} \left ( \frac{H[M_{\rm{lim}}(z_i)-M_k] }{ \sum_j^{N_p} \Phi_j \Delta M H[M_{\rm{lim}}(z_i) - M_j]} \right )}\,.
 \end{equation}
 
 This estimator is then applied iteratively, each time replacing the input values for $\Phi_j$ with the output $\Phi_k$ values from the previous iteration until sufficient convergence is achieved.   
 
 \subsection{Comparison of Four LF Estimators}
 \label{lfcomp}

\begin{figure}
\begin{center}
\includegraphics[width=80mm]{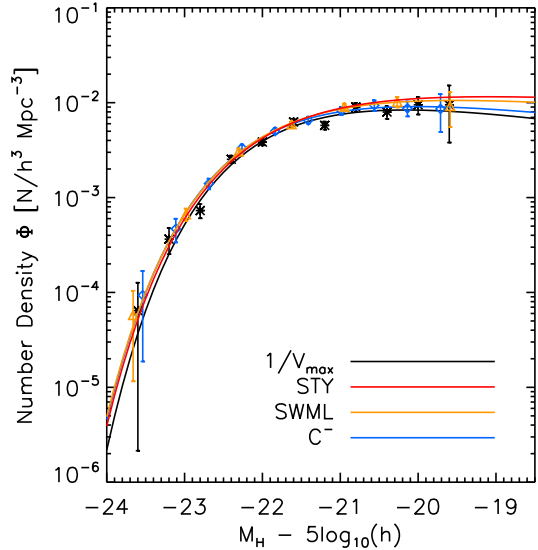}
\caption{\label{lfcompare} A comparison of the $H-$band LFs derived for our sample using four different methods.  The only method that establishes the normalization is the $1/V_{max}$ method, so all four methods have been set to that normalization for comparison.  }
\end{center}
\end{figure} 

\begin{figure}
\begin{center}
\includegraphics[width=80mm]{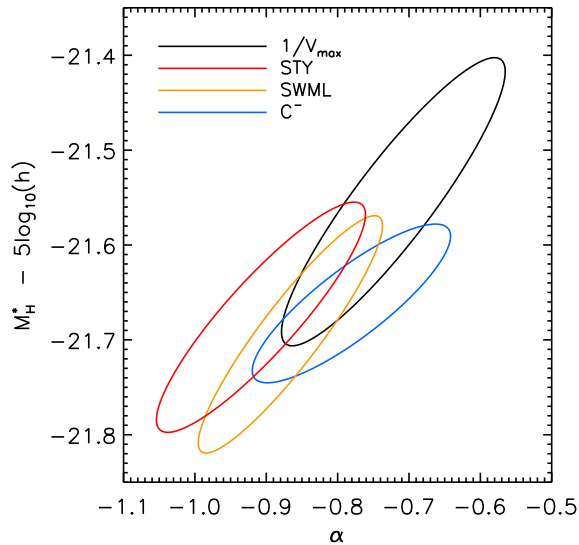}
\caption{\label{macompare} A comparison of the $H-$band LF parameters $M^*$ and $\alpha$ derived for our sample using four different methods.  The ellipses show the $1~\sigma$ errors in these parameters for each method.   }
\end{center}
\end{figure}

In Figures~\ref{lfcompare}~and~\ref{macompare}, we compare the LFs and Schechter function parameters $M^*$ and $\alpha$ determined from the four LF estimators described in Sections~\ref{parlf}~and~\ref{nonparlf} ($1/V_{max}$, STY, SWML, and $C^-$) .  

Figure~\ref{lfcompare}  shows a comparison of the $H-$band LF derived for our sample using the four different methods.  The $1/V_{max}$ method is shown in black, the STY method in red, the $C^-$ method in blue, and the SWML method in orange.  The only method that establishes the normalization is the $1/V_{max}$ method, so all four methods have been set to that normalization for comparison.   

Figure~\ref{macompare} shows a comparison of the $H-$band LF parameters $M^*$ and $\alpha$ derived for our sample using the four different methods.  The same color designations apply as for Figure~\ref{lfcompare}.    

We expect the $\alpha$ parameter for our work to be close to the value of $\alpha \approx -1$ found in large low-redshift studies, as evolution in the LF should be minimal since $z \sim 0.2$.  Given our comparison of different methods above, we decided to use the STY method to determine the $\alpha$ parameter in the sections that follow.  We determine the error in our calculation of $\alpha$ by fitting a 2-dimensional Gaussian to the distribution of values for the likelihood function in $\alpha, M^*$ space.  We find that a 2-d Gaussian is a good fit to the distribution.  We show the $1~\sigma$ error ellipse of this Gaussian is that shown in Figure~\ref{macompare}.   

In further exploration of fitting $\alpha$ with STY, we found that the method is sensitive to the input apparent magnitude limit of the survey.  Our sample is $H-$band selected, such that the magnitude limit is well known in the $H-$band but less well defined in the $J~$and$~K-$bands.  Thus, we first determine      $\alpha$ in the $H-$band ($\alpha=0.91 \pm 0.11$) and then hold that value fixed in our determination of $M^*$ using STY for $J, H$ and $K$.  

In Section~\ref{cv}, we discuss a comparison with the COSMOS cone mock catalogs to study cosmic variance. We use the mock catalogs to test our method of determining $\alpha$ and $M^*$ with STY.  To do this we combined all the mock catalogs to form a simulated survey 24 times the size of our survey.  We applied the above methods to determine $\alpha$ and $M^*$.  We then created mock surveys of 6 random sightlines the same size as our survey and performed the same process to test whether the values for $\alpha$ and $M^*$ recovered from the smaller survey were consistent with the global values.  We found that the recovered values for $\alpha$ and $M^*$ for the smaller mock surveys were in $1~\sigma$ agreement with the global values determined from combining all the mock catalogs.  

\subsection{Determining The LF Normalization}
Of the four methods discussed above for determining the LF Schechter function parameters, the $1/V_{max}$ method is the only one that gives the normalization directly.  The other methods require that the normalization be determined by other means.  

Several methods have been developed for determining the LF normalization, and these are studied in detail by \citet{Will97} via simulated data.  They find that the various density estimators yield roughly equivalent results, and all tend to underestimate the true density of objects by some $\sim 20\%$.  The parameter $\phi^*$ is related to the mean density ($\overline{n}$) of the sample through

\begin{equation}
\phi^* = \frac{\overline{n}}{\int_{M_{bright}}^{M_{faint}} \Phi (M) dM}\,,
\end{equation}

\noindent where $M_{bright}$~and~$M_{faint}$ are the absolute magnitude limits of the survey.

One commonly used method for determining the mean density is the minimum variance estimator of \citet{Davi82}

\begin{equation}
n_{minvar}=\frac{\sum_{i=1}^{N_{gal}} w(z_i)}{\int dV S(z) w(z)}, 
\end{equation}

\noindent where $N_{gal}$ is the number of galaxies in the sample, $S(z)$ is the selection function, and $w(z)$ is a weighting function for each galaxy defined by the inverse of the second moment ($J_3$) of the two-point correlation function, $\xi(r)$.  

The selection function, $S(z)$, is defined as

\begin{equation}
S(z) = \frac{\int_{M_{bright}}^{min(M_{max(z_i)}, M_{faint})} \phi (M) dM}{\int_{M_{bright}}^{M_{faint}} \phi (M) dM}\,, 
\end{equation}

\noindent where $M_{max(z_i)}$ is the faintest absolute magnitude detectable at redshift $z_i$.  

The weighting function is defined as

\begin{equation}
w(z) = \frac{1}{1+ \overline{n} J_3 S(z)} \,, ~~~J_3 = \int_0^r r^2 \xi (r) dr \,.
\end{equation}

In the case that all weights are set to a value of 1, the minimum variance estimator described above reduces to the so called ``$n_3$'' estimator proposed by \citet{Davi82}

\begin{equation}
n_3 = \frac{N_T}{\int_0^{z_{max}} S(z) dV} \,.
\end{equation}

Another estimator proposed by \citet{Davi82} is

\begin{equation}
n_1=  \left [ \int_0^{z_{max}} \frac{N(z)}{S(z)} dz  \right ] \bigg / \left [  \int_0^{z_{max}} \frac{dV}{dz} dz \right ],
\end{equation}

\noindent which in the case of 1 galaxy per bin reduces to the form used by \citet{Efst88}:

\begin{equation}
n_1 = \frac{1}{V} \sum_{i=1}^{N_{gal}} \frac{1}{S(z_i)} \,.
\end{equation}

We show a comparison of the normalization derived using the aforementioned mean density estimators of \citet{Davi82} alongside the result for the $1/V_{max}$ method in Table~\ref{lftable2}.  In all cases, we have determined $\alpha$ and $M^*$ using the STY method.  We find that all four estimates of the LF normalization agree to within 1~$\sigma$ statistical errors.   

\input{Table2}

\subsection{A Hybrid Method for Determining the NIR LF}

All LFs for our observed data presented in the following sections are calculated using the method described in Section~\ref{lfcomp}, where, using STY, we first determine $\alpha$ from the $H-$band sample and then $M^*$ for each bandpass.  We then determine $\phi^*$ by fitting a Schechter function to the $1/V_{max}$ results in each bandpass with $\alpha$ and $M^*$ fixed at the values determined with the STY method.  

The $1/V_{max}$ method requires binning. Thus, we compared results from bin sizes ranging from $0.2 - 1$ magnitudes.  We found the results to be consistent to within $1~\sigma$ (statistical uncertainty in the resulting Schechter function fit).    We settled on a bin size of $0.4$ magnitudes to provide the best counting statistics over a wide range in absolute magnitudes.   

The STY method does not allow for any test of goodness of fit between data and model.  We fit the Schechter function parameter $\phi^*$ to the $1/V_{max}$ binned data, with $\alpha$ and $M^*$ fixed, using the \emph{mpfit.pro} least-squares fitting procedure of \citet{Mark09}, which is based on the \emph{MINPACK}\footnote{http://www.netlib.org/minpack} algorithm of Jorge Mor{\'e}.  The resulting fits yield a $\chi^2 \sim 15.5$ for 10 degrees of freedom, suggesting that the Schechter function is a good fit to the data at the $\sim 90\%$ confidence level.  In Section~\ref{cv}, we discuss the systematic uncertainties in our determination of the LF normalization due to cosmic variance.

To calculate absolute magnitudes we modified the observed apparent magnitude by a distance modulus $(DM)$, a $K-$correction $K(z)$, and an evolution correction $E(z)$, such that $M=m-DM(z)-K(z) + E(z)$.

\subsection{NIR $K-$corrections $K(z)$}
At low redshifts, $K-$corrections are small and nearly independent of galaxy type in the NIR \citep{Mann01}.  Using SDSS and UKIDSS data, \citet{Chil10} showed that, at low redshifts ($z<0.5$), accurate $K-$corrections may be calculated using low order polynomials and inputs of only redshift and one observed NIR color.  They provide a $K-$correction calculator\footnote{http://kcor.sai.msu.ru/}, which we used to compute $K-$corrections for galaxies in our sample given redshifts and NIR colors.  

We find the calculator output values to be in agreement with the expected NIR $K-$corrections for galaxies at these redshifts derived by \citet{Mann01}.   Given the analysis of \citet{Chil10} in comparing their calculator outputs with more rigorous SED fitting methods for determining $K-$corrections, we expect our magnitude errors associated with the $K-$corrections should be $<\pm 0.1$.  
 
 \subsection{Evolution Corrections $E(z)$}
 Evolution in the rest-frame NIR for  galaxies at relatively low redshifts is expected to be weak but still has a significant effect on the normalization of the LFs, particularly at $z>0.1$.  The simplest  form of the evolution correction is $E(z)=Qz$ (e.g., \citealt{Smit09}), where $Q$ is a positive constant.  \citet{Blan03} showed that in the NIR, $Q=1$ agrees well with stellar population synthesis models.  Thus, for this work we adopt $Q_J = Q_H = Q_K = 1$, such that $E(z) = z$ in all bands.  
  
   
\subsection{The Correlation Between Schechter Function Parameters}
\begin{figure}
\begin{center}
\includegraphics[width=80mm]{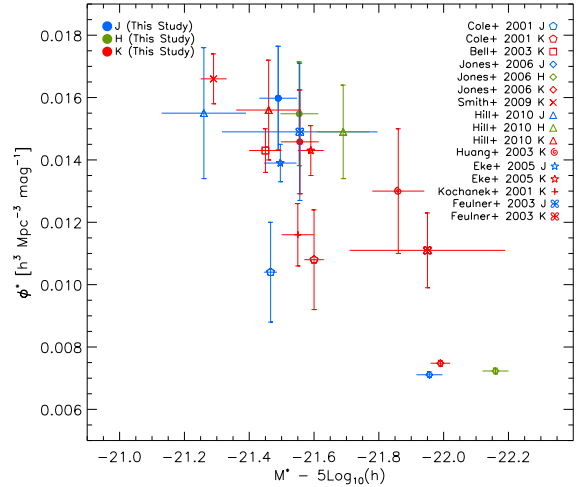}
\caption{\label{mstarvphi} $\phi^*$ versus $M^*$ for various studies.  Studies in the $J-$band are shown in blue, $H-$band in green, and $K-$band in red.  The error bars show the $1~\sigma$ statistical errors in each study.  We note the trend toward a lower value for $\phi^*$ at brighter values for $M^*$. }
\end{center}
\end{figure} 

\begin{figure}
\begin{center}
\includegraphics[width=80mm]{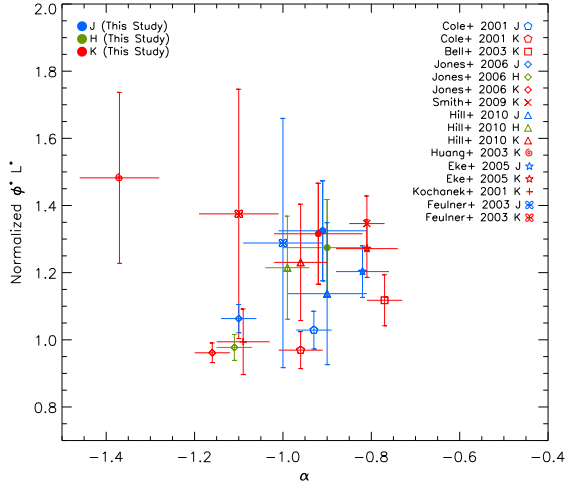}
\caption{\label{alphavphil} The normalized product $\phi^* L^*$ versus $\alpha$ for various studies.  Studies in the $J-$band are shown in blue, $H-$band in green, and $K-$band in red.  The error bars show the $1~\sigma$ statistical errors in each study }
\end{center}
\end{figure} 

The Schechter function parameters $\phi^*$ and $M^*$ are correlated, as shown in Figure~\ref{mstarvphi}.  Thus, rather than a direct comparison of $\phi^*$ values from various studies, we consider the product of the normalization and the characteristic luminosity ($\phi^* L^*$) in the following discussions of comparisons between our study and those from the literature.  
  
The Schechter function parameter $\alpha$ is also correlated with $\phi^*$ and $M^*$ but uncorrelated with the product of the normalization and characteristic luminosity, as shown in Figure~\ref{alphavphil}.  Thus, our comparison of the product $\phi^* L^*$ between studies is unaffected by the variations in $\alpha$ from one study to another.  

\subsection{Comparing Studies With Different Photometric Methods}
\label{photocomp}
Here we briefly discuss the photometric methods employed in different studies that we compare with in this work.   For all our photometry, we calibrate to 2MASS point sources and use the SExtractor \citep{BA96} MAG\_AUTO aperture, which fits a Kron-like \citep{Kron80} ellipse to each galaxy.  In \citet{Keen10a} we found that this aperture does an excellent job of retrieving total magnitudes for galaxies at high signal to noise (galaxies in the current study are $4-5$ magnitudes above our detection limits of \citealt {Keen10a}).   

In the case of \citet{Koch01}, \citet{Cole01}, \citet{Bell03}, \citet{Eke05},  and \citet{Jone06}, 2MASS Kron or similar apertures were used for photometry, and we expect this to be consistent with our work.  

 \citet{Hill10}  apply Kron aperture photometry to the UKIDSS Large Area Survey (LAS).  \citet{Smit09} also work with the UKIDSS LAS, but they use Petrosian magnitudes.  They compare the difference between Petrosian and Kron magnitudes for $\sim 7,000$ galaxies that also have photometry in 2MASS.   They find that  2MASS Kron magnitudes are, in general, $\sim 0.2$ magnitudes brighter than their UKIDSS Petrosian counterparts.  We adjust the value for $M^*$ from \citet{Smit09} to accommodate this difference.  
 
\citet{Huan03} use circular $8\arcsec$ apertures corrected to $20\arcsec$ apertures for galaxies at $K_{\rm{Vega}}>13$.  While it is difficult to assess exactly how this method compares to our photometry, the errors on their measurement are large and so this does not affect our analysis.  Similarly, the sample size of \citet{Feul03} is small enough that the associated uncertainties span essentially the entire range of measurements of $\phi^* L^*$ from the literature.

 \subsection{The Effects of Magnitude Errors on the LFs}
The fundamental selection for this work comes from $H-$band photometry we obtained with the Ultra-Low Background Camera (ULBCam, \citealt{Hall04}) on the University of Hawaii 2.2 m telescope.  Our final $H-$band mosaics are highly uniform and $4-5$ magnitudes deeper than the selection limit for this work ($H < 18$).  We calibrated our photometry for all six fields using high signal to noise point sources in 2MASS and found no discernable systematics and a random magnitude error of $< 0.05$.  For the galaxies in this survey, which are extended and somewhat fainter than the calibration sources, our random magnitude error is $\sim 0.1$.

\subsection{The Effects of Redshift Errors on the LFs}
Our redshift sample consists of $\sim 8\%$ photometric redshifts and $\sim 8\%$ questionable spectroscopic redshifts.  As demonstrated by comparison with the SDSS, the high-confidence spectroscopic sample could be considered accurate to within $\Delta z \sim 0.0005$.  Thus, here we consider only the error associated with questionable and photometric redshifts.

If we assume, as an upper limit, that all questionable and photometric redshifts have an associated random error of $\Delta z/(1+z) \sim 0.1$, then this would imply a random error in a calculated absolute magnitude of $\sim 1-1.5$.  Although a large effect for any individual galaxy, the contribution of this effect to the mean magnitude error in any given magnitude bin would be of the same order as the contribution from the random magnitude errors associated with the spectroscopic sample.  

\subsection{The Effects of Evolution Corrections on the LFs}
\citet{Cole01} showed that for NIR LFs at $\langle z \rangle \sim 0.05$, the evolution correction has the effect of raising the normalization by $\sim 10\%$ and dimming $M^*$ by $\sim 0.1$ magnitudes (though these changes were of the same order as their errors).  \citet{Jone06} did not apply an evolution correction for their sample at $\langle z \rangle \sim 0.05$, and they measure a significantly lower normalization and brighter $M^*$ than \citet{Cole01}.   In general, however, applying an evolution correction does not significantly change the product of the normalization and characteristic luminosity ($\phi^* L^*$).  

\subsection{Effects of Magnitude Errors in General}
The $E(z)$ and $K(z)$ errors, along with the aforementioned magnitude errors, amount to a total error in the calculated absolute magnitudes of $\sim 0.3$.  If these errors are truly random, then the effect on the LFs will be to shift galaxies from one bin to another, but the randomness of the process should essentially cancel itself out.

\subsection{The NIR Luminosity Function at $0.1<z<0.3$}
\begin{figure*}
\begin{center}
\includegraphics[width=150mm]{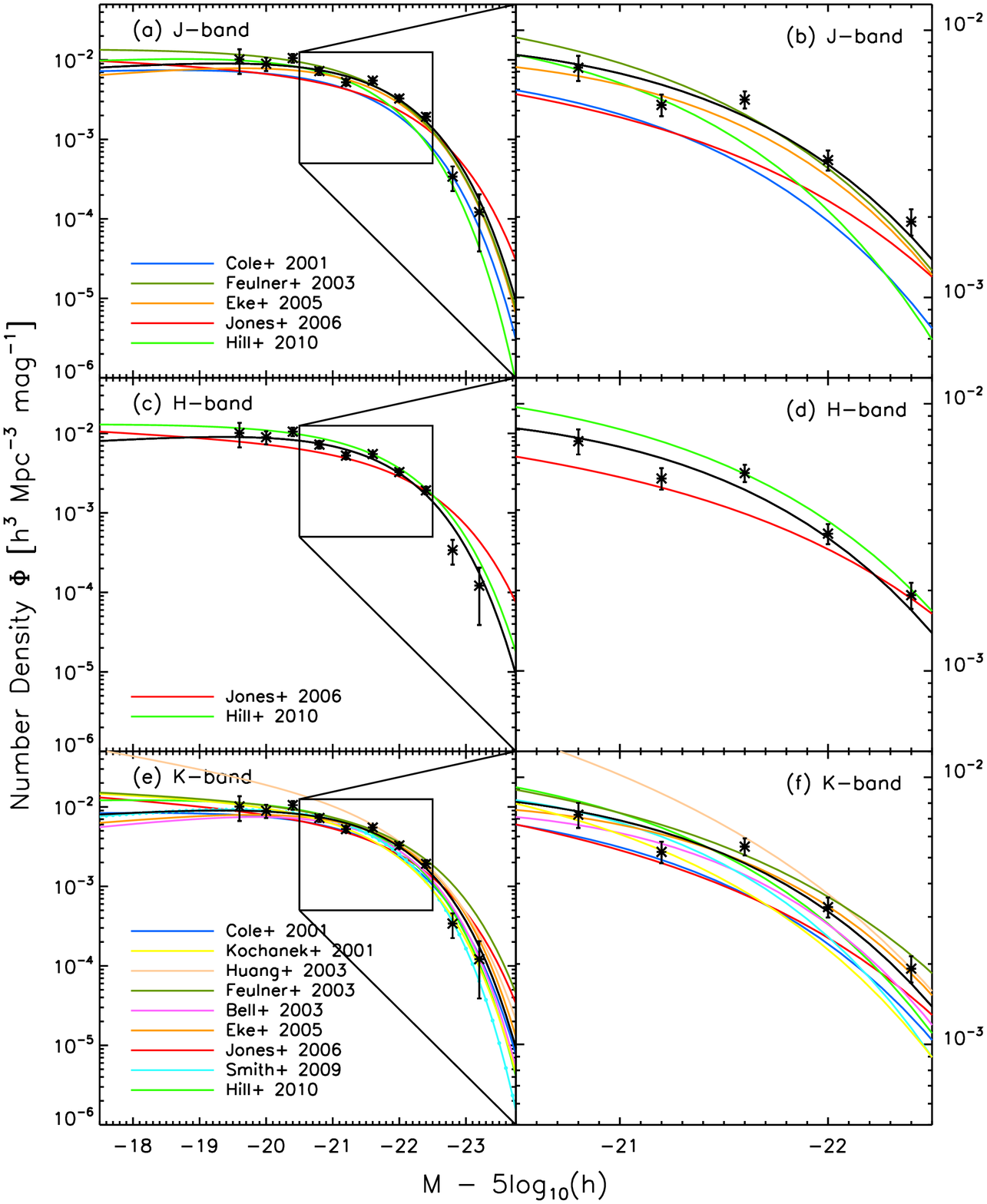}
\caption{\label{lf} The $J, H,$~and~$K-$band LFs of galaxies in our six fields combined for the redshift range $0.1<z<0.3$.  We derived the parameters $M^*$ and $\alpha$ via the STY method and then applied the $1/V_{max}$ method and fit a Schechter function with $M^*$ and $\alpha$ fixed to derive the normalization $\phi^*$.  In the left hand panels (a, c, and e) we plot the entire LFs to show overall shape for comparison with lower redshift LFs from the literature.  The colored lines show LFs from the studies denoted in the plots.  In the right hand panels, we zoom in on the normalization to highlight the differences between the normalization in this study and selections from the literature.  The solid lines show the Schechter fits to the combined data for all six fields (five fields in the $K-$band where SSA17 lacks photometry).  The error bars shown are the 1$~\sigma$ Poisson errors taken from \citet{Gehr86}.  The normalizations and $M^*$ values to the fitted Schechter functions  for the combination of all six fields are given in Table~\ref{lftable3}.}
\end{center}
\end{figure*} 

\begin{figure*}
\begin{center}
\includegraphics[width=180mm]{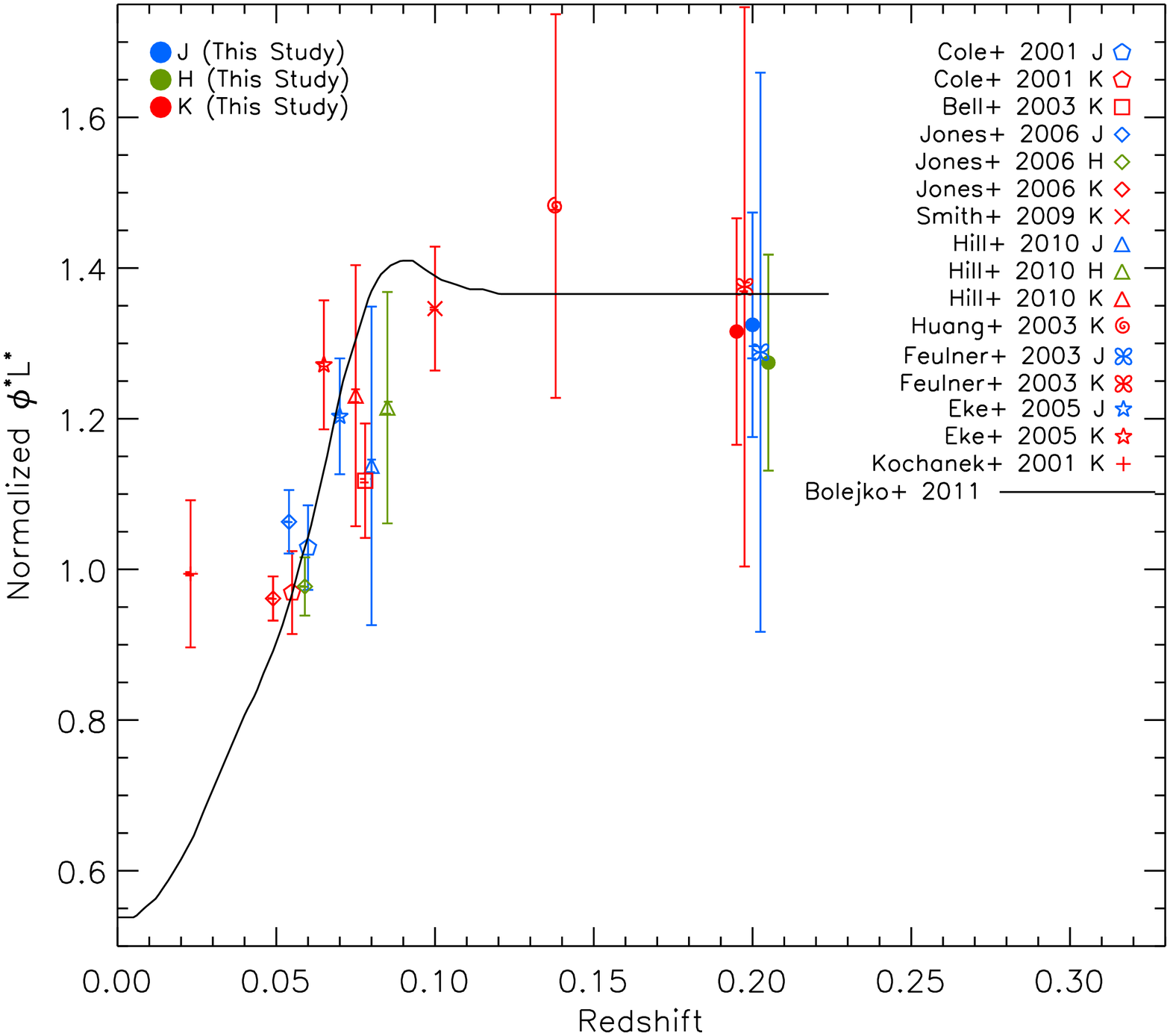}
\caption{\label{zvphim} The product of the Schechter function normalization and characteristic luminosity ($\phi^* L^*$) for NIR LFs in this study (filled circles) and selections from the literature.  We have employed a color correction, such that all data from $J, H,$ and $K$ appear vertically on the same scale.  We then normalized the $\phi^* L^*$ values to an error weighted average of the large studies at $\langle z \rangle = 0.057$ of \citet{Cole01} and \citet{Jone06}.  All measurements from our study and from the literature for the $J-$band are shown in blue, $H-$band in green, and $K-$band in red.   In each case, $\phi^* L^*$ is plotted at the median redshift for galaxies in a given study. Where studies (including this one) have covered more than one bandpass, the $J-$band $\phi^* L^*$ is plotted at the median redshift of the study and $H$ and/or $K-$band values are shifted slightly along the abscissa for clarity.    The error bars show the 1 $\sigma$ statistical errors in the product $\phi^* L^*$.  The values for $\phi^*$ and $L^* (M^*)$ used to generate this figure are given in Table~\ref{lftable3}. The solid line shows the void radial density profile from \citet{Bole11a} that they claim would be sufficient to provide an apparent acceleration of the expansion of the universe observed via type Ia supernovae.  \citet{Bole11a} presented this void profile as a density contrast; $\langle \rho(r) \rangle / \rho_0$.  We have converted from their density contrast to $\phi^* L^*$ by normalizing to the results of \citet{Cole01} and \citet{Jone06}. }
\end{center}
\end{figure*} 

\input{Table3}

In Figure~\ref{lf}, we show the $J, H,$~and~$K-$band LFs of galaxies in our six fields combined for the redshift range $0.1<z<0.3$.  In the left-hand panels (a, c, and e), the entire LFs are shown to demonstrate  the overall shape compared with lower redshift selections from the literature.  In the right-hand panels, we zoom in on the normalization ($M^* \sim -21.7$) to highlight the differences between this study and other studies from the literature.  The solid lines show the Schechter fits to the combined data for all six fields (five fields in the $K-$band where SSA17 lacks photometry).  

Given the $H-$band value for $M^* \sim -21.7$, we can see from Figure~\ref{zhist} that we are sampling $M \sim M^*$ galaxies out to a redshift of $z\sim 0.3$.  We also note from the distribution shown in Figure~\ref{zhist} that we don't have a significant number of $M \sim M^*$ galaxies below a redshift of $z\sim 0.1$, such that the normalization of the NIR LF is best sampled in this survey for $0.1<z<0.3$.

In Figure~\ref{zvphim}, we show the product of the LF normalization and characteristic luminosity ($\phi^* L^*$) as a function of redshift from this study (filled circles) and from some literature studies.  All measurements in the $J-$band are shown in blue, $H-$band in green, and $K-$band in red.   We applied a color correction so that all three bandpasses would appear on the same scale vertically.  We then normalized the $\phi^* L^*$ values to an error-weighted average of the large studies at $\langle z \rangle = 0.057$ of \citet{Cole01} and \citet{Jone06}.  In each case, $\phi^* L^*$ is plotted at the median redshift for galaxies in a given study.  Where studies (including this one) have covered more than one bandpass, the $J-$band $\phi^* L^*$ is plotted at the median redshift of the study, and $H$ and/or $K-$band values are shifted slightly along the abscissa for clarity.   The error bars show the 1 $\sigma$ statistical errors in $\phi^*$ for each study.  

The solid black line in Figure~\ref{zvphim} shows the void radial density profile from \citet{Bole11a}, which they claim could produce the apparent acceleration of the expansion of the universe observed via type Ia supernovae.  \citet{Bole11a} present this void profile as a density contrast profile, $\langle \rho(r) \rangle /\rho_0$.  We convert their density contrast profile to normalized $\phi^* L^*$ by normalizing to the average of the large low-redshift studies of  \citet{Cole01} and \citet{Jone06}.  We note that such a radial density profile is not ruled out by current measurements of the NIR LF normalization.

We find that our measured value of the product $\phi^* L^*$ is $\sim 30\%$ higher than the mean value of studies at $\langle z \rangle \sim 0.05$.  Our measurement could be considered a conservative underestimate of the true value of $\phi^* L^*$ for $0.1 < z < 0.3$, because our survey avoids known galaxy clusters in this redshift range.  
 



\section{Cosmic Variance}
 \label{cv}
 
The relative scatter in measurements of the NIR LF normalization over the past ten years could be arising from a variety of sources, including differences in fitting or normalization methods, differences in $K(z)$ or $E(z)$ corrections, photometry errors or incomplete spectroscopy, and cosmic variance.  Most likely, these differences arise from some combination of the aforementioned sources, but in the majority of modern surveys, cosmic variance accounts for a large fraction of the uncertainty budget.  Thus, we wish to quantify and compare the effects of cosmic variance on the uncertainties in our measurements and those from the literature.  First, we consider uncertainties due to cosmic variance in all recent measurements of the LF normalization in the literature.

\subsection{Cosmic Variance in Recent Measurements of the NIR LF normalization}

\citet{Driv10} have quantified cosmic variance for $M^*$ galaxies in the SDSS volume.  They derive the following empirical formula to calculate the expected systematics due to cosmic variance in any low-redshift survey

\begin{eqnarray}
\rm{Cosmic~Variance} (\%) = [1.00-0.03\sqrt{(A/B)-1}] \nonumber \\
\times (219.7-52.4 \rm{log}_{10}[AB \times 291.0] \nonumber \\
+ 3.21(\rm{log}_{10}[AB \times 291.0])^2)/\sqrt{NC/291.0}
\end{eqnarray}

In the above formula, A and B are the transverse comoving lengths at the median redshift of the survey, and thus provide an aspect ratio $A/B$ for rectangular surveys ($\rm{A}>\rm{B}$).  C is the comoving length of the redshift interval surveyed, and N is the number of independent sightlines composing the survey.  Given this formula, it is clear that surveys with multiple sightlines suffer less from cosmic variance systematics (by a factor of $1/\sqrt{N}$) than single field surveys, and surveys with large aspect ratios are superior to those of square fields.  

We calculated the expected systematics due to cosmic variance for each survey from the literature, given depth, area, aspect ratio, and number of fields as inputs to the empirical formula of \citet{Driv10}.  We were not able to account for the fact that, in many cases, these surveys consisted of several sightlines, each having a different size and aspect ratio, so these estimates are not meant to be extremely rigorous, but rather to provide a means of comparing the relative effects of cosmic variance from one survey to the next.  The results of this comparison are listed in Table~\ref{lftable3}.  Below, we briefly discuss this estimate of cosmic variance in each survey from the literature.

The first efforts using large surveys to measure the local NIR LF normalization were \citet{Cole01} and \citet{Koch01}.  Both of these surveys used 2MASS for their NIR photometry, and \citet{Cole01} used 2dFGRS spectroscopy for $5,683$ galaxies at a median redshift $z \sim 0.06$, while \citet{Koch01} used the ZCAT data from the CfA2 redshift survey \citep{Gell89,Huch92} for $3,878$ galaxies at a median redshift $z \sim 0.023$.  We find that both these surveys should suffer from systematics due to cosmic variance at the $\sim 10\%$ level.  \citet{Cole01} estimate cosmic variance systematics at the $15\%$ level, using galaxy counts from 2MASS,  and propagate this estimate through their error estimation.  

\citet{Eke05} provide an updated version of the study done by \citet{Cole01} using 2MASS and the completed 2dFGRS to include $15, 664$ galaxies in the determination of the $K-$band LFs.  Given the larger survey size, the estimated systematics due to cosmic variance in this sample are at the $\sim 6\%$ level. 

The largest study to date of the local NIR LF was done by \citet{Jone06}.  They combine 2MASS photometry with the 6-degree Field Galaxy Redshift Survey (6dFGS, \citealt{Jone04}) to measure the $J,H,$~and~$K-$band LFs using $\sim 60,000$ galaxies at a median redshift $z\sim 0.054$.  While this study includes more galaxies than previous works, the fact that it is relatively shallow implies that cosmic variance still persists at the $5\%$ level.  

At median redshifts $z\sim 0.08$, \citet{Bell03} and \citet{Hill10} provide a measurement of the LF normalization.  \citet{Bell03} use a sample $6,282$ galaxies selected from 2MASS with spectroscopy from the SDSS early data release.  They find a normalization quite similar to that of \citet{Eke05}, though with larger error bars (cosmic variance at the $\sim 10\%$ level) due to the smaller sample size.  \citet{Hill10} combine the UKIDSS LAS, the SDSS, and the Millennium Galaxy Catalog (MGC, \citealt{Lisk03}) to form a highly uniform and $100\%$ spectroscopically complete sample of $\sim 1800$ galaxies in the $J,H,$~and~$K-$bands.  Their LF normalization agrees well with that of \citet{Bell03}, though the relatively small volume of their sample implies possible systematics on the $\sim 30\%$ level due to cosmic variance.  

\citet{Smit09} use a sample of $\sim 40,000$ galaxies at a median redshift $z\sim 0.1$, drawn from the UKIDSS LAS with spectroscopy from the SDSS, to study the $K-$band LF.   Their sample includes all UKIDSS LAS galaxies at $K_{\rm{Vega}}<16$ that have secure redshifts in the SDSS. While this is not a spectroscopically complete magnitude-limited sample, it is one of the largest samples used to study the $K-$band LF and samples a large volume.  For this survey size, we estimate cosmic variance to be at the $2\%$.  However, given the relative spectroscopic incompleteness, other biases may be a significant source of systematic uncertainties.

At median redshifts in the range $0.1<z<0.3$, very few studies of the NIR LF have been done.  Furthermore, the relatively small sample sizes imply that measurements of the LF normalization over this range are subject to large systematic uncertainties due to cosmic variance.   \citet{Huan03} use a sample of 1,056 galaxies to study the $K-$band LF at a median redshift of $z \sim 0.14$.  Their measured normalization is essentially in agreement with all measurements at lower redshift, given their statistical errors plus cosmic variance at the $\sim 10\%$ level.    \citet{Feul03} use a sample of 210 galaxies to study the $J$ and $K-$band LF at a median redshift of $z \sim 0.2$.  The size of their survey  implies uncertainties due to cosmic variance at the $\sim 30\%$ level, putting their measured normalization in agreement with all previous studies.  

Our survey samples the same redshift range as \citet{Feul03}, but our larger area and deeper completeness limits mean we have a factor of four more galaxies in our study, and the expected uncertainties due to cosmic variance have been reduced by a factor of two.  As noted above, our measured normalization is in good agreement with studies at $z\sim 0.1$, but would appear to be in slight tension with studies at $z \sim 0.05$.  In the following section we explore the field to field variability in the LF normalization for our survey, and compare with simulations using the COSMOS Cone mock catalogs. 

\subsection{Cosmic Variance in Our Survey}

\input{Table4}

\begin{figure}
\begin{center}
\includegraphics[width=80mm]{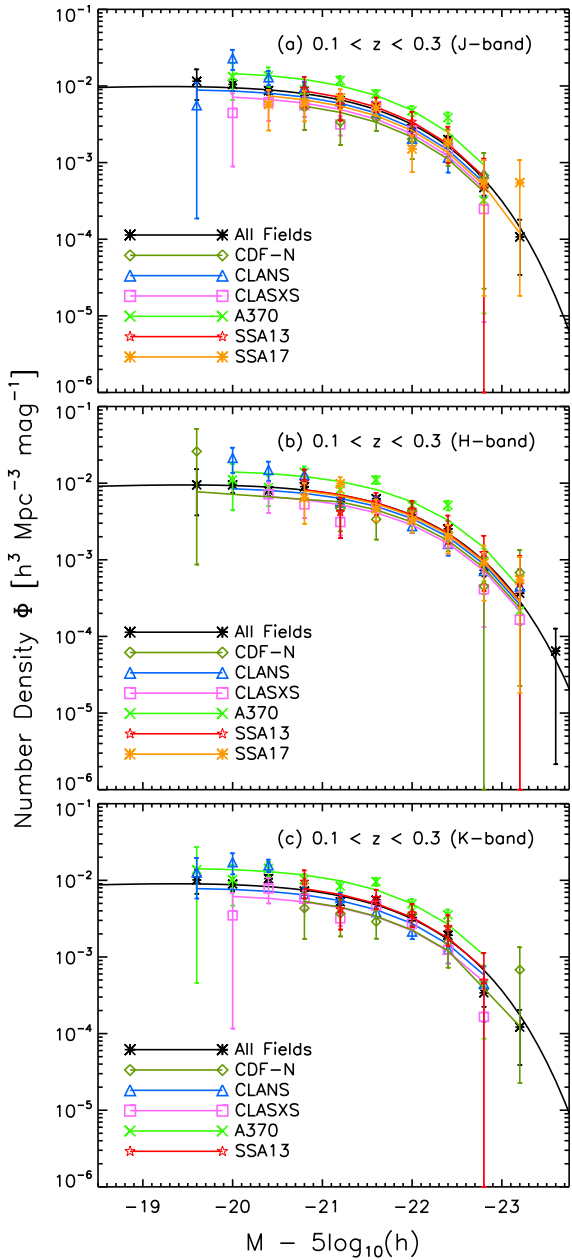}
\caption{\label{f2flf} The $J, H,$~and~$K-$band LFs of galaxies  for $0.1 < z < 0.3$ in the $J$(a),~$H$(b),~and~$K$(c) bands.  The solid black line shows Schechter fits to the combination of all six fields (black asterisks, only five fields in the $K-$band where SSA17 lacks photometry). The colored lines show Schechter fits for each of the six fields individually.  These LFs are derived in the same way as those presented previously, with Schechter fits derived from the results of the $1/V_{max}$ method using a fixed $\alpha$ and $M^*$ derived from the STY method.  Error bars show the $1~\sigma$ Poisson errors from \citet{Gehr86}.  The Schechter function normalizations for each individual field in this redshift selection are given in Table~\ref{lftable4}.  We note that the normalization in the A370 field is roughly a factor of two higher than in the other fields, even though the A370 cluster itself, at a redshift of $z=0.37$ is excluded.}
\end{center}
\end{figure}

We are measuring the normalization of the NIR LF over the redshift range $0.1 < z < 0.3$, or comoving distances of $\sim 300-850~h^{-1}$~Mpc.  Thus, with a total of 2~deg$^2$ on the sky we are sampling the normalization over a volume of $\sim10^5~h^{-3}~$Mpc$^3$.  With typical large-scale structures existing on at least $\sim 100~h^{-1}$~Mpc scales, we expect (and indeed find) significant field to field variation in the measured normalization.   

In Figure~\ref{f2flf}, we show the $J, H,$~and~$K-$band LFs for $0.1 < z < 0.3$ on a field to field basis.  In all cases the A370 field shows a much higher normalization than the other five fields, even though the A370 cluster itself, at redshift $z=0.37$, is not included in these LFs.  The solid black line shows Schechter fits to the combination of all six fields. The colored lines show Schechter fits for each of the six fields individually.

In Table~\ref{lftable4} we give the normalizations for each field and the combination of fields in the range $0.1<z<0.3$.  These values were calculated using the $1/V_{max}$ method and fitting a Schechter function with fixed $\alpha = 0.91$ from applying STY in the $H-$band, and fixed $M^*$ values (given in Table~\ref{lftable3} derived in each bandpass using STY with a fixed $\alpha = 0.91$).

We find agreement between the normalizations in all fields except the A370 field, where the normalization is a factor of $\sim 2$ higher.  Upon investigation of this apparent excess, we find that two sheet-like overdensities exist in the A370 field at redshifts of $z\sim0.18$~and~$z\sim0.25$~(both clearly visible in Figure~\ref{zhist}).  These overdensities are very narrow in redshift space, spanning the redshift ranges $0.17<z<0.19$~and~$0.245<z<0.255$, respectively.  To determine whether our survey appears to be consistent with cosmological simulations, and to further explore the expected systematics due to cosmic variance in this study, we use the COSMOS Cone mock catalogs.

\subsection{Comparison With The COSMOS Cones}

\begin{figure}
\begin{center}
\includegraphics[width=90mm]{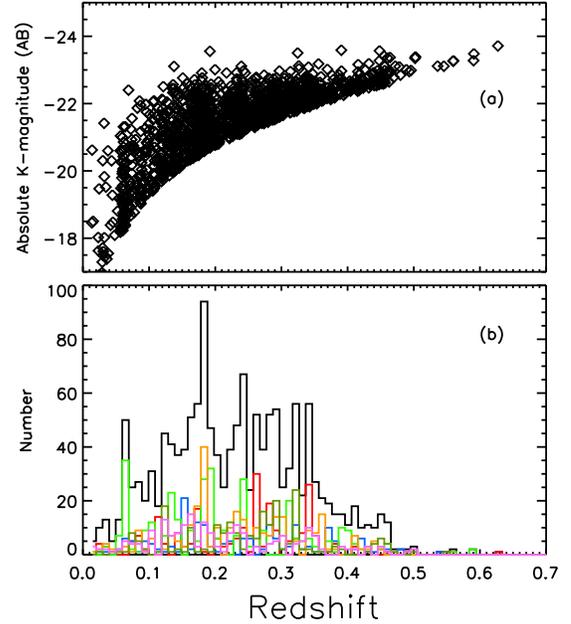}
\caption{\label{zhistcosmo}  (a) Redshift versus $K-$band absolute magnitude in an example selection of galaxies ($K<18$) for the combination of six sightlines of 1/3~deg$^2$ each taken at random from the COSMOS cone mock catalogs.  (b) A histogram of all redshifts in the six mock catalogs.  The solid black line shows the total over six fields and the colored lines represent the six fields individually.  We note the similarity in the distribution of redshift and absolute magnitudes in these example mock catalogs in comparison to our observations shown in Figure~\ref{zhist}.}
\end{center}
\end{figure} 

\begin{figure}
\begin{center}
\includegraphics[width=80mm]{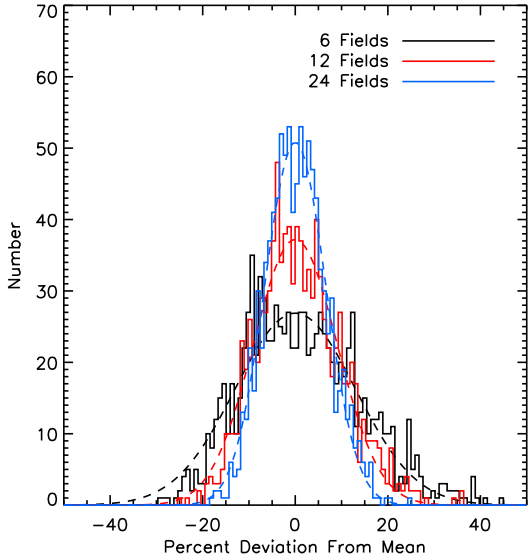}
\caption{\label{gauss}  Percent deviation from the mean NIR LF normalization for simulated surveys of 6 (black), 12 (red), and 24 (blue) sightlines.   The histograms show the distribution in measured normalization for 1000 realizations of the simulated surveys from the COSMOS cone mock catalogs.   The dashed lines show a Gaussian fit to each histogram. Each survey is intended to mimic our observed data, in that the simulated data are magnitude limited ($K<18$) and restricted to the redshift range $0.1<z<0.3$. }
\end{center}
\end{figure} 

\begin{figure}
\begin{center}
\includegraphics[width=80mm]{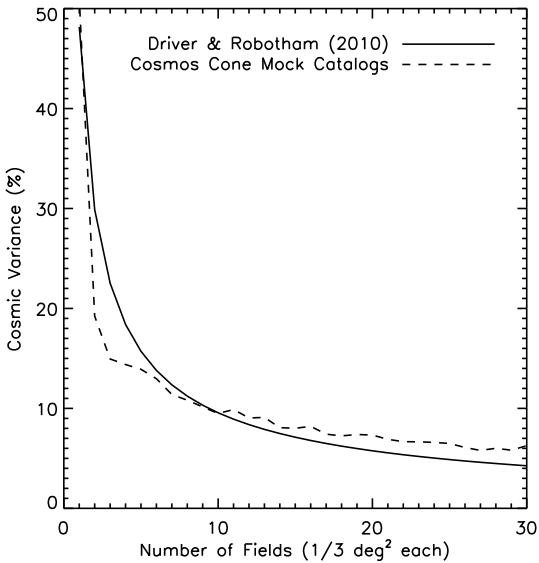}
\caption{\label{cosvar}  Systematic uncertainty in percent (1~$\sigma$) due to cosmic variance as a function of the number of sightlines in surveys composed of multiple square sightlines of $1/3~$deg$^2$ each ($K<18$ and $0.1<z<0.3$. The solid line shows the empirical estimate from the formula of \citet{Driv10} and the dashed line shows our estimate calculated from the COSMOS Cone mock catalogs.}
\end{center}
\end{figure}

The COSMOS cones  are composed of 24 pencil beam mock catalogs created by Manfred Kitzbichler \citep{Kitz07} for the Cosmic Evolution Survey collaboration (COSMOS, \citealt{Scov07a}) and made public in the Millennium Database online\footnote{http://gavo.mpa-garching.mpg.de/MyMillennium3/Help?page=databases/mpamocks/cosmos2006}.   These catalogs were created from the Millenium simulation \citep{Spri05} using the galaxy formation algorithm of \citet{Delu06} and the merger tree prescriptions of \citet{Delu07}.  The COSMOS cone catalogs have been used primarily to study galaxy formation and evolution, as well as large-scale structure, at relatively high redshifts \citep{Scov07b, Mccr07, Knob09, Kova10}.  We use the catalogs here to compare with our observations of the NIR LFs at $0.1<z<0.3$.  

The COSMOS cone mock catalogs represent 8 sightlines from each of 3 different origins within the Millenium simulation.  Each of the COSMOS cone pencil beams covers an area of $1.4 \times 1.4$ degrees, or $\sim 2$~deg$^2$.  Thus, each mock catalog is roughly the equivalent area of our entire survey, and so we divide each mock catalog into six separate sightlines.  This gives us a total of 144 mock sightlines to compare with our observations.  $K-$band apparent magnitudes are provided in the mock catalogs (though not $J$~or~$H-$band), so we take a magnitude limited sample from the mocks of $K<18$ to compare with our data.  

From the 144 sightlines, we first select six at random. We show an example of the distribution in absolute $K-$band magnitude and redshift for six random sightlines selected from the COSMOS cones in Figure~\ref{zhistcosmo}.  This apparent magnitude selection of $K<18$ typically generated a catalog of $\sim 1500$ objects in the combination of six sightlines.  We note the similarity in the distribution of redshift and absolute magnitude in comparison to our observed data for $H<18$ shown in Figure~\ref{zhist}.

Next, we construct LFs to study the variability in the normalization from field to field.  Rather than using the absolute magnitudes provided in the catalogs, we attempted to mimic our methodology applied to the observed data by using the apparent $K-$band magnitudes from the mock catalogs, and converting to absolute magnitude via a distance modulus, a $K-$correction from \citet{Mann01}, and an $E(z)=z$ correction.  We then compared the normalization of the LFs in six randomly selected mock catalogs in the same manner as with the observational data, where we estimated $\alpha$ and $M^*$ with the STY method, then found the normalization fitting a Schechter function to the $1/V_{max}$ results with a fixed $\alpha$ and $M^*$.  

We completed 1000 realizations of the process of selecting six sightlines at random and evaluating their LFs.   We found that we have a $\sim 25\%$ chance of drawing a survey like ours, with one overdense sightline, from the mock catalogs.  Thus, while the appearance of overdensities like those in the A370 field might not be characteristic of sightlines chosen at random from simulations, they are not strongly atypical.  However, we note that in both observations and simulations, massive galaxy clusters tend to lie at the confluence of large filamentary structures, such that choosing to observe in the direction of a galaxy cluster may increase the chances of finding overdensities along the line of sight.

Next, we measured the expected systematic uncertainty due to cosmic variance by looking at the distribution of measured normalizations for our $1,000$ mock surveys of six sightlines.   We find the distribution is well fit by a Gaussian (see Figure~\ref{gauss}) with a $1~\sigma$ variability of $\sim 13\%$ in the measured normalization over the $1,000$ combinations of six random sightlines.  

We repeated the process of 1000 realizations of simulated surveys with the number of sightlines (1/3 deg$^2$ each) ranging from $1-30$.  Also shown in Figure~\ref{gauss} is the distribution in measured normalization for surveys of 12 and 24 sightlines for comparison.  
   
In Figure~\ref{cosvar} we show the $1~\sigma$ systematic uncertainty (in percent) due to cosmic variance as a function of the number of square 1/3 deg$^2$ sightlines.  The dashed line shows our results from simulated surveys drawn from the COSMOS cones.  The expected systematics  given the empirical formula of \citet{Driv10} are shown as a solid line.  We find that the two estimates agree rather well, and hence we confirm their results derived from observed data with our analysis of the COSMOS cones.  Given these results, we find that doubling the number of sightlines in our survey should reduce the effects of cosmic variance to the $\sim 10\%$ level, while quadrupling the survey should reduce these systematics to the $\sim 5\%$ level.

\section{Summary}
\label{summary}
An accurate measurement of the average space density of galaxies just beyond the local universe ($z>0.1$) is of interest, in particular, because recent cosmological modeling efforts have shown that if the matter density at $z<0.1$ is low by roughly $50\%$, then the apparent acceleration of the expansion of the universe observed in type Ia supernovae could simply be a product of our location with respect to local structure  \citep{Alex09, Bole11a}.  

We have presented a study of the NIR LF normalization at redshifts of $ 0.1 < z < 0.3$.  This study is based on a highly complete ($> 90\%$) spectroscopic sample of 812 galaxies, selected to be brighter than 18$^{th}$ magnitude in the $H-$band, over six widely-separated fields at high Galactic latitudes.  While a relatively small sample of galaxies, our survey represents a factor of four increase in the number of galaxies in previous studies of the NIR LF in this redshift range, which implies a factor of two decrease in the systematic uncertainties due to cosmic variance.

We construct the NIR LFs for each field individually and for the combination of all six fields.  We compare the product of the normalization and characteristic luminosity ($\phi^* L^*$) in the combination of our six fields with lower redshift measurements to test for a higher luminosity density at $z>0.1$.  We find our measurement of $\phi^* L^*$  to be in agreement with other studies at median redshifts near $z \sim 0.1$, but roughly $30\%$ higher than the error-weighted mean  at $\langle z \rangle \sim 0.05$.    Our measurement of $\phi^* L^*$ for $0.1 < z < 0.3$ could be considered a conservative underestimate of the true value of $\phi^* L^*$ because we avoid known galaxy clusters in this redshift range.  

We use the COSMOS cone mock catalogs, and the empirical formula of \citet{Driv10} to investigate the effects of cosmic variance on our measurement of the NIR LF normalization and on those from the literature.  Despite the uncertainties due to cosmic variance, it would appear that data from the literature are beginning to suggest a rising luminosity density from $z=0.05$ to $z=0.1$.  While this conclusion is tentative at best, it would not be terribly surprising, as a rising LF normalization over this redshift range would be predicted given the results of several NIR galaxy counts surveys over the past $10-20$ years.  

More importantly, however, we require knowledge of the luminosity density at $z>0.1$ to understand whether or not the local universe is under-dense.  We find  that measurements of the NIR LF at $z>0.1$, including our own, are still too uncertain to provide a robust comparison with lower redshift measurements.  While inconclusive, this result is noteworthy given the possible implications for locally measured cosmological observables.  

Surveys that are currently underway, including our own campaign of spectroscopic follow-up on fields that have existing NIR photometry, will soon provide a more robust answer to the question of whether or not we reside in a large local underdensity.  

\acknowledgements{We thank the anonymous referee for a careful review of this article and the comments and suggestions that helped to improve the manuscript.

We gratefully acknowledge support from the NSF grants
AST 0708793 (A.~J.~B.) and AST 0709356
(L.~L.~C.), the University of Wisconsin Research Committee with funds granted
by the Wisconsin Alumni Research Foundation, and the David and Lucile Packard
Foundation (A.~J.~B.).  R.~C.~K. was supported by a
Wisconsin Space Grant Consortium Graduate Fellowship, a Sigma Xi Grant in
Aid of Research, an NSF East Asia and Pacific Summer Institutes Fellowship, and a Fulbright Fellowship during various portions of this work. 

We thank Dr. Ralf Kotulla for support and suggestions using \emph{GALEV} and \emph{GAZELLE}, and for providing the SED models from his dissertation work for our use in this study.

This research made use of the ``K-corrections calculator'' service available at http://kcor.sai.msu.ru/.

This work is based in part on observations obtained with WIRCam, a joint
project of CFHT, Taiwan, Korea, Canada, France, and the Canada-France-Hawaii
Telescope (CFHT) which is operated by the National Research Council (NRC) of
Canada, the Institute National des Sciences de l'Univers of the Centre
National de la Recherche Scientifique of France, and the University of
Hawaii.  

The Millennium Simulation databases used in this paper and the web application providing online access to them were constructed as part of the activities of the German Astrophysical Virtual Observatory.

This publication makes use of data products from the Sloan Digital Sky Survey (SDSS). Funding for the SDSS and SDSS-II has been provided by the Alfred P. Sloan Foundation, the Participating Institutions, the National Science Foundation, the U.S. Department of Energy, the National Aeronautics and Space Administration, the Japanese Monbukagakusho, the Max Planck Society, and the Higher Education Funding Council for England. The SDSS Web Site is http://www.sdss.org/. The SDSS is managed by the Astrophysical Research Consortium for the Participating Institutions. The Participating Institutions are the American Museum of Natural History, Astrophysical Institute Potsdam, University of Basel, University of Cambridge, Case Western Reserve University, University of Chicago, Drexel University, Fermilab, the Institute for Advanced Study, the Japan Participation Group, Johns Hopkins University, the Joint Institute for Nuclear Astrophysics, the Kavli Institute for Particle Astrophysics and Cosmology, the Korean Scientist Group, the Chinese Academy of Sciences (LAMOST), Los Alamos National Laboratory, the Max-Planck-Institute for Astronomy (MPIA), the Max-Planck-Institute for Astrophysics (MPA), New Mexico State University, Ohio State University, University of Pittsburgh, University of Portsmouth, Princeton University, the United States Naval Observatory, and the University of Washington.}  

This work has made use of NASA's Astrophysics Data System.


\end{document}

%% file: Table1.tex
\begin{deluxetable*}{lcccccc}
\tabletypesize{\tiny}
\tablewidth{0pt}
\tablecaption{\label{summarytab}Coordinates, areas, and number of targets
  (spectroscopic redshifts / total)}
\tablehead{\textbf{Field} & \textbf{CDF-N} & \textbf{CLANS}  & \textbf{CLASXS}
  &\textbf{SSA13} &\textbf{SSA17} &\textbf{A370}} \\
\startdata
R.A.(hh:mm:ss) & 12:36:55  & 10:46:54 & 10:34:58 & 13:12:16 & 17:06:31 & 02:39:53\\
Dec (dd:mm:ss) & 62:14:19  & 59:08:26 & 57:52:22 & 42:41:24 & 43:55:44 &-01:34:37\\ 
Galactic l (deg) &125.9  &148.2 &151.5 &109.1 & 68.9 & 173\\
Galactic b (deg) &54.8   &51.4  &51.0  &73.8 & 42.0 & -53.5\\
Supergalactic l (deg) &54.7   &52.2  &52.3  & 75.3 & 65.1 & 302.3\\
Supergalactic b (deg) &11.7  &-1.8  &-3.9  & 13.6 & 49.5 & -25.7\\
Area (deg$^2$) &0.2 & 0.5 & 0.5 & 0.2 & 0.2 & 0.4\\
Spectroscopic Redshifts &92 & 297 & 278 & 121 & 161 & 367\\
Total Targets ($H<18$) &100 & 313 & 298 & 124 & 188 & 413\\

\enddata
\end{deluxetable*}

%% file: Table2.tex
\begin{deluxetable}{lc}
\tabletypesize{\tiny}
\tablewidth{0pt}
\tablecaption{\label{lftable2}$H-$band NIR LF normalization
$\phi^*$\tablenotemark{a} derived using four different methods\tablenotemark{b}}
\tablehead{ \textbf{Method} ~~~~~~~~~~~~~~~~~~~~~~& \textbf{$\phi^* \times 10^3$}} 
\startdata

$1/V_{max}$ & $15.3 \pm 1.6$ \\
$n_{minvar}$ & $15.6 \pm 1.4$ \\
$n_1$ & $14.9 \pm 1.5$ \\
$n_3$ & $15.7 \pm 1.3$ \\

\enddata
\tablenotetext{a}{all $\phi^*$ values are given in units of $h^3$~Mpc$^{-3}$.}
\tablenotetext{b}{The four methods listed are $1/V_{max}$, plus three methods ($n_{minvar}, n_1, n_3$) used by \citet{Davi82}. }

\end{deluxetable}

%% file: Table3.tex
\begin{deluxetable*}{lcccccccc}
\tabletypesize{\tiny}
\tablewidth{0pt}
\tablecaption{\label{lftable3}NIR luminosity functions parameters $\phi^*$\tablenotemark{a} and $M^*$\tablenotemark{b} for this study and selections from the literature}
\tablehead{ \textbf{Study} && \multicolumn{2}{c}{\textbf{$J-$band}} & \multicolumn{2}{c}{\textbf{$H-$band}} & \multicolumn{2}{c}{\textbf{$K-$band}}} \\
\startdata
 & $\langle z \rangle$  & $\phi^* \times 10^3$ & $M^*$ & $\phi^* \times 10^3$ & $M^*$ & $\phi^* \times 10^3$ & $M^*$ &CV\tablenotemark{c}\\
This Study\tablenotemark{d} & 0.2 &15.9$\pm$1.7 &-21.49$\pm$0.05 &15.3$\pm$1.6& -21.67$\pm$0.05&14.6$\pm$1.7&-21.56$\pm$0.06&14\%\\
Feulner et al. 2003 & 0.2 &14.9$\pm$2.2&-21.56$\pm$0.24&$-$&$-$&11.1$\pm$1.2&-21.95$\pm$0.24&27\%\\
Huang et al. 2003 & 0.14 &$-$&$-$&$-$&$-$&13.0$\pm$2.0&-21.86$\pm$0.08&11\%\\
Smith\tablenotemark{e} et al. 2009 & 0.1 &$-$&$-$&$-$&$-$&16.6$\pm$0.8&-21.49$\pm$0.05&2\%\\
Bell et al. 2003 & 0.08 &$-$&$-$&$-$&$-$&14.3$\pm$0.7&-21.45$\pm$0.05&10\%\\
Hill et al. 2010 & 0.08 &15.5$\pm$2.1&-21.46$\pm$0.13&14.9$\pm$1.5&-21.89$\pm$0.08&15.6$\pm$1.6&-21.66$\pm$0.10&30\%\\
Eke et al. 2005 & 0.07 &13.9$\pm$0.6&-21.50$\pm$0.05&$-$&$-$&14.3$\pm$0.8&-21.59$\pm$0.04&6\%\\
Cole et al. 2001 & 0.06 &10.4$\pm$1.6&-21.47$\pm$0.02&$-$&$-$&10.8$\pm$1.6&-21.60$\pm$0.02&8\%\\
Jones et al. 2006 & 0.05 &7.1$\pm$0.1&-21.96$\pm$0.04&7.2$\pm$0.1&-22.16$\pm$0.04&7.5$\pm$0.1&-21.99$\pm$0.03&5\%\\
Kochanek et al. 2001 & 0.023 &$-$&$-$&$-$&$-$&11.6$\pm$1.0&-21.55$\pm$0.05&11\%\\

\enddata
\tablenotetext{a}{all $\phi^*$ values are given in units of $h^3$~Mpc$^{-3}$.  For this study we use a fixed $\alpha = -0.91$ derived in the $H-$band using the STY method as described in Section~\ref{lfcomp}.}
\tablenotetext{b}{all $M^*$ values are given as $M-5\rm{log}_{10}(h)$}
\tablenotetext{c}{Cosmic variance estimates in percent using the empirical formula of \citet{Driv10}}
\tablenotetext{d}{The errors in $\phi^*$ represent the errors in the Schechter fits plus the systematics due to cosmic variance.}
\tablenotetext{e}{The $M^*$ value listed for Smith et al. is adjusted to be $0.2$ magnitudes brighter than their published value to accommodate the difference between Petrosian and Kron apertures as described in Section~\ref{photocomp}.}
\end{deluxetable*}

%% file: Table4.tex
\begin{deluxetable}{lccc}
\tabletypesize{\scriptsize}
\tablewidth{0pt}
\tablecaption{\label{lftable4}NIR Luminosity Function Normalizations\tablenotemark{a} in each of six\tablenotemark{b} fields for the redshift range $0.1 < z < 0.3$}
\tablehead{ \textbf{Field} & $\phi_J^* \times 10^3$ & $\phi_H^* \times 10^3$ & $\phi_K^* \times 10^3$ }
\startdata
All Fields &15.9$\pm$1.7&15.3$\pm$1.6&14.6$\pm$1.7\\
A370 &24.4$\pm$4.9&23.1$\pm$4.7&23.0$\pm$4.5\\
CLANS &14.4$\pm$3.5&13.9$\pm$3.4&12.7$\pm$3.1\\
CLASXS &12.0$\pm$3.3&11.5$\pm$3.2&10.3$\pm$2.8\\
CDF &10.9$\pm$6.4&12.5$\pm$7.4&10.4$\pm$6.3\\
SSA13 &17.3$\pm$7.7&15.5$\pm$7.7&15.3$\pm$7.6\\
SSA17 &13.3$\pm$5.6&15.0$\pm$6.2& N/A \\
\enddata
\tablenotetext{a}{all $\phi^*$ values are given in units of $h^3$~Mpc$^{-3}$.  These values were calculated using the $1/V_{max}$ method and fitting a Schechter function with fixed $\alpha = 0.91$ from applying STY in the $H-$band, and fixed $M^*$ values (given in Table~\ref{lftable3} derived in each bandpass using STY with a fixed $\alpha = 0.91$.  This method is described in Section~\ref{lfcomp}}
\tablenotetext{b}{only five fields in the case of the $K-$band where the SSA17 field lacks photometry}
\end{deluxetable}

%% file: Keen0512.bbl
\begin{thebibliography}{103}
\expandafter\ifx\csname natexlab\endcsname\relax\def\natexlab#1{#1}\fi

\bibitem[{{Alexander} {et~al.}(2003){Alexander}, {Bauer}, {Brandt},
  {Schneider}, {Hornschemeier}, {Vignali}, {Barger}, {Broos}, {Cowie},
  {Garmire}, {Townsley}, {Bautz}, {Chartas}, \& {Sargent}}]{Alex03}
{Alexander}, D.~M., {et~al.} 2003, \aj, 126, 539

\bibitem[{{Alexander} {et~al.}(2009){Alexander}, {Biswas}, {Notari}, \&
  {Vaid}}]{Alex09}
{Alexander}, S., {Biswas}, T., {Notari}, A., \& {Vaid}, D. 2009, JCAP, 9, 25

\bibitem[{{Alnes} {et~al.}(2006){Alnes}, {Amarzguioui}, \& {Gr{\o}n}}]{Alne06}
{Alnes}, H., {Amarzguioui}, M., \& {Gr{\o}n}, {\O}. 2006, \prd, 73, 083519

\bibitem[{{Barden} \& {Armandroff}(1995)}]{Bard95}
{Barden}, S.~C., \& {Armandroff}, T. 1995, in Presented at the Society of
  Photo-Optical Instrumentation Engineers (SPIE) Conference, Vol. 2476, Society
  of Photo-Optical Instrumentation Engineers (SPIE) Conference Series, ed.
  {S.~C.~Barden}, 56--67

\bibitem[{{Barger} {et~al.}(2008){Barger}, {Cowie}, \& {Wang}}]{Barg08}
{Barger}, A.~J., {Cowie}, L.~L., \& {Wang}, W.~H. 2008, \apj, 689, 687

\bibitem[{{Barro} {et~al.}(2009){Barro}, {Gallego}, {P{\'e}rez-Gonz{\'a}lez},
  {Eliche-Moral}, {Balcells}, {Villar}, {Cardiel}, {Cristobal-Hornillos}, {Gil
  de Paz}, {Guzm{\'a}n}, {Pell{\'o}}, {Prieto}, \& {Zamorano}}]{Barr09}
{Barro}, G., {et~al.} 2009, \aap, 494, 63

\bibitem[{{Bell} \& {de Jong}(2001)}]{Bell01}
{Bell}, E.~F., \& {de Jong}, R.~S. 2001, \apj, 550, 212

\bibitem[{{Bell} {et~al.}(2003){Bell}, {McIntosh}, {Katz}, \&
  {Weinberg}}]{Bell03}
{Bell}, E.~F., {McIntosh}, D.~H., {Katz}, N., \& {Weinberg}, M.~D. 2003, \apjs,
  149, 289

\bibitem[{{Bershady} {et~al.}(2008){Bershady}, {Barden}, {Blanche}, {Blanco},
  {Corson}, {Crawford}, {Glaspey}, {Habraken}, {Jacoby}, {Keyes}, {Knezek},
  {Lemaire}, {Liang}, {McDougall}, {Poczulp}, {Sawyer}, {Westfall}, \&
  {Willmarth}}]{Bers08}
{Bershady}, M., {et~al.} 2008, in Presented at the Society of Photo-Optical
  Instrumentation Engineers (SPIE) Conference, Vol. 7014, Society of
  Photo-Optical Instrumentation Engineers (SPIE) Conference Series

\bibitem[{{Bertin} \& {Arnouts}(1996)}]{BA96}
{Bertin}, E., \& {Arnouts}, S. 1996, \aaps, 117, 393

\bibitem[{{Biswas} {et~al.}(2010){Biswas}, {Notari}, \& {Valkenburg}}]{Bisw10}
{Biswas}, T., {Notari}, A., \& {Valkenburg}, W. 2010, JCAP, 11, 30

\bibitem[{{Blanton} {et~al.}(2003){Blanton}, {Hogg}, {Bahcall}, {Brinkmann},
  {Britton}, {Connolly}, {Csabai}, {Fukugita}, {Loveday}, {Meiksin}, {Munn},
  {Nichol}, {Okamura}, {Quinn}, {Schneider}, {Shimasaku}, {Strauss}, {Tegmark},
  {Vogeley}, \& {Weinberg}}]{Blan03}
{Blanton}, M.~R., {et~al.} 2003, \apj, 592, 819

\bibitem[{{Bolejko} {et~al.}(2011){Bolejko}, {C{\'e}l{\'e}rier}, \&
  {Krasi{\'n}ski}}]{Bole11b}
{Bolejko}, K., {C{\'e}l{\'e}rier}, M.-N., \& {Krasi{\'n}ski}, A. 2011,
  Classical and Quantum Gravity, 28, 164002

\bibitem[{{Bolejko} \& {Sussman}(2011)}]{Bole11a}
{Bolejko}, K., \& {Sussman}, R.~A. 2011, Physics Letters B, 697, 265

\bibitem[{{Brandt} {et~al.}(2001){Brandt}, {Alexander}, {Hornschemeier},
  {Garmire}, {Schneider}, {Barger}, {Bauer}, {Broos}, {Cowie}, {Townsley},
  {Burrows}, {Chartas}, {Feigelson}, {Griffiths}, {Nousek}, \&
  {Sargent}}]{Bran01}
{Brandt}, W.~N., {et~al.} 2001, \aj, 122, 2810

\bibitem[{{Bruzual} \& {Charlot}(2003)}]{Bruz03}
{Bruzual}, G., \& {Charlot}, S. 2003, \mnras, 344, 1000

\bibitem[{{Bull} \& {Clifton}(2012)}]{Bull12}
{Bull}, P., \& {Clifton}, T. 2012, ArXiv e-prints

\bibitem[{{Busswell} {et~al.}(2004){Busswell}, {Shanks}, {Frith}, {Outram},
  {Metcalfe}, \& {Fong}}]{Buss04}
{Busswell}, G.~S., {Shanks}, T., {Frith}, W.~J., {Outram}, P.~J., {Metcalfe},
  N., \& {Fong}, R. 2004, \mnras, 354, 991

\bibitem[{{C{\'e}l{\'e}rier} {et~al.}(2010){C{\'e}l{\'e}rier}, {Bolejko}, \&
  {Krasi{\'n}ski}}]{Cele10}
{C{\'e}l{\'e}rier}, M.-N., {Bolejko}, K., \& {Krasi{\'n}ski}, A. 2010, \aap,
  518, A21

\bibitem[{{Chilingarian} {et~al.}(2010){Chilingarian}, {Melchior}, \&
  {Zolotukhin}}]{Chil10}
{Chilingarian}, I.~V., {Melchior}, A.-L., \& {Zolotukhin}, I.~Y. 2010, \mnras,
  405, 1409

\bibitem[{{Chung} \& {Romano}(2006)}]{Chun06}
{Chung}, D.~J.~H., \& {Romano}, A.~E. 2006, \prd, 74, 103507

\bibitem[{{Clarkson} \& {Maartens}(2010)}]{Clar10}
{Clarkson}, C., \& {Maartens}, R. 2010, Classical and Quantum Gravity, 27,
  124008

\bibitem[{{Cole} {et~al.}(2001){Cole}, {Norberg}, {Baugh}, {Frenk},
  {Bland-Hawthorn}, {Bridges}, {Cannon}, {Colless}, {Collins}, {Couch},
  {Cross}, {Dalton}, {De Propris}, {Driver}, {Efstathiou}, {Ellis},
  {Glazebrook}, {Jackson}, {Lahav}, {Lewis}, {Lumsden}, {Maddox}, {Madgwick},
  {Peacock}, {Peterson}, {Sutherland}, \& {Taylor}}]{Cole01}
{Cole}, S., {et~al.} 2001, \mnras, 326, 255

\bibitem[{{Colless} {et~al.}(2001){Colless}, {Dalton}, {Maddox}, {Sutherland},
  {Norberg}, {Cole}, {Bland-Hawthorn}, {Bridges}, {Cannon}, {Collins}, {Couch},
  {Cross}, {Deeley}, {De Propris}, {Driver}, {Efstathiou}, {Ellis}, {Frenk},
  {Glazebrook}, {Jackson}, {Lahav}, {Lewis}, {Lumsden}, {Madgwick}, {Peacock},
  {Peterson}, {Price}, {Seaborne}, \& {Taylor}}]{Coll01}
{Colless}, M., {et~al.} 2001, \mnras, 328, 1039

\bibitem[{{Cowie} {et~al.}(2004){Cowie}, {Barger}, {Fomalont}, \&
  {Capak}}]{Cowi04}
{Cowie}, L.~L., {Barger}, A.~J., {Fomalont}, E.~B., \& {Capak}, P. 2004, \apjl,
  603, L69

\bibitem[{{Davis} \& {Huchra}(1982)}]{Davi82}
{Davis}, M., \& {Huchra}, J. 1982, \apj, 254, 437

\bibitem[{{Davis} {et~al.}(2007){Davis}, {Guhathakurta}, {Konidaris}, {Newman},
  {Ashby}, {Biggs}, {Barmby}, {Bundy}, {Chapman}, {Coil}, {Conselice},
  {Cooper}, {Croton}, {Eisenhardt}, {Ellis}, {Faber}, {Fang}, {Fazio},
  {Georgakakis}, {Gerke}, {Goss}, {Gwyn}, {Harker}, {Hopkins}, {Huang},
  {Ivison}, {Kassin}, {Kirby}, {Koekemoer}, {Koo}, {Laird}, {Le Floc'h}, {Lin},
  {Lotz}, {Marshall}, {Martin}, {Metevier}, {Moustakas}, {Nandra}, {Noeske},
  {Papovich}, {Phillips}, {Rich}, {Rieke}, {Rigopoulou}, {Salim},
  {Schiminovich}, {Simard}, {Smail}, {Small}, {Weiner}, {Willmer}, {Willner},
  {Wilson}, {Wright}, \& {Yan}}]{Davi07}
{Davis}, M., {et~al.} 2007, \apjl, 660, L1

\bibitem[{{de Jong}(1996)}]{Dejo96}
{de Jong}, R.~S. 1996, \aap, 313, 377

\bibitem[{{De Lucia} \& {Blaizot}(2007)}]{Delu07}
{De Lucia}, G., \& {Blaizot}, J. 2007, \mnras, 375, 2

\bibitem[{{De Lucia} {et~al.}(2006){De Lucia}, {Springel}, {White}, {Croton},
  \& {Kauffmann}}]{Delu06}
{De Lucia}, G., {Springel}, V., {White}, S.~D.~M., {Croton}, D., \&
  {Kauffmann}, G. 2006, \mnras, 366, 499

\bibitem[{{Driver} \& {Robotham}(2010)}]{Driv10}
{Driver}, S.~P., \& {Robotham}, A.~S.~G. 2010, \mnras, 407, 2131

\bibitem[{{Driver} {et~al.}(2011){Driver}, {Hill}, {Kelvin}, {Robotham},
  {Liske}, {Norberg}, {Baldry}, {Bamford}, {Hopkins}, {Loveday}, {Peacock},
  {Andrae}, {Bland-Hawthorn}, {Brough}, {Brown}, {Cameron}, {Ching}, {Colless},
  {Conselice}, {Croom}, {Cross}, {de Propris}, {Dye}, {Drinkwater}, {Ellis},
  {Graham}, {Grootes}, {Gunawardhana}, {Jones}, {van Kampen}, {Maraston},
  {Nichol}, {Parkinson}, {Phillipps}, {Pimbblet}, {Popescu}, {Prescott},
  {Roseboom}, {Sadler}, {Sansom}, {Sharp}, {Smith}, {Taylor}, {Thomas},
  {Tuffs}, {Wijesinghe}, {Dunne}, {Frenk}, {Jarvis}, {Madore}, {Meyer},
  {Seibert}, {Staveley-Smith}, {Sutherland}, \& {Warren}}]{Driv11}
{Driver}, S.~P., {et~al.} 2011, \mnras, 413, 971

\bibitem[{{Efstathiou} {et~al.}(1988){Efstathiou}, {Ellis}, \&
  {Peterson}}]{Efst88}
{Efstathiou}, G., {Ellis}, R.~S., \& {Peterson}, B.~A. 1988, \mnras, 232, 431

\bibitem[{{Eke} {et~al.}(2005){Eke}, {Baugh}, {Cole}, {Frenk}, {King}, \&
  {Peacock}}]{Eke05}
{Eke}, V.~R., {Baugh}, C.~M., {Cole}, S., {Frenk}, C.~S., {King}, H.~M., \&
  {Peacock}, J.~A. 2005, \mnras, 362, 1233

\bibitem[{{Enqvist} \& {Mattsson}(2007)}]{Enqv07}
{Enqvist}, K., \& {Mattsson}, T. 2007, JCAP, 2, 19

\bibitem[{{Fazio} {et~al.}(2004){Fazio}, {Hora}, {Allen}, {Ashby}, {Barmby},
  {Deutsch}, {Huang}, {Kleiner}, {Marengo}, {Megeath}, {Melnick}, {Pahre},
  {Patten}, {Polizotti}, {Smith}, {Taylor}, {Wang}, {Willner}, {Hoffmann},
  {Pipher}, {Forrest}, {McMurty}, {McCreight}, {McKelvey}, {McMurray}, {Koch},
  {Moseley}, {Arendt}, {Mentzell}, {Marx}, {Losch}, {Mayman}, {Eichhorn},
  {Krebs}, {Jhabvala}, {Gezari}, {Fixsen}, {Flores}, {Shakoorzadeh}, {Jungo},
  {Hakun}, {Workman}, {Karpati}, {Kichak}, {Whitley}, {Mann}, {Tollestrup},
  {Eisenhardt}, {Stern}, {Gorjian}, {Bhattacharya}, {Carey}, {Nelson},
  {Glaccum}, {Lacy}, {Lowrance}, {Laine}, {Reach}, {Stauffer}, {Surace},
  {Wilson}, {Wright}, {Hoffman}, {Domingo}, \& {Cohen}}]{Fazi04}
{Fazio}, G.~G., {et~al.} 2004, \apjs, 154, 10

\bibitem[{{February} {et~al.}(2010){February}, {Larena}, {Smith}, \&
  {Clarkson}}]{Febr10}
{February}, S., {Larena}, J., {Smith}, M., \& {Clarkson}, C. 2010, \mnras, 405,
  2231

\bibitem[{{Feulner} {et~al.}(2003){Feulner}, {Bender}, {Drory}, {Hopp},
  {Snigula}, \& {Hill}}]{Feul03}
{Feulner}, G., {Bender}, R., {Drory}, N., {Hopp}, U., {Snigula}, J., \& {Hill},
  G.~J. 2003, \mnras, 342, 605

\bibitem[{{Frith} {et~al.}(2003){Frith}, {Busswell}, {Fong}, {Metcalfe}, \&
  {Shanks}}]{Frit03}
{Frith}, W.~J., {Busswell}, G.~S., {Fong}, R., {Metcalfe}, N., \& {Shanks}, T.
  2003, \mnras, 345, 1049

\bibitem[{{Frith} {et~al.}(2005){Frith}, {Shanks}, \& {Outram}}]{Frit05}
{Frith}, W.~J., {Shanks}, T., \& {Outram}, P.~J. 2005, \mnras, 361, 701

\bibitem[{{Garcia-Bellido} \& {Haugb{\o}lle}(2008)}]{Garc08a}
{Garcia-Bellido}, J., \& {Haugb{\o}lle}, T. 2008, JCAP, 4, 3

\bibitem[{{Garc{\'{\i}}a-Bellido} \& {Haugb{\o}lle}(2008)}]{Garc08b}
{Garc{\'{\i}}a-Bellido}, J., \& {Haugb{\o}lle}, T. 2008, JCAP, 9, 16

\bibitem[{{Garc{\'{\i}}a-Bellido} \& {Haugb{\o}lle}(2009)}]{Garc09}
---. 2009, JCAP, 9, 28

\bibitem[{{Gehrels}(1986)}]{Gehr86}
{Gehrels}, N. 1986, \apj, 303, 336

\bibitem[{{Geller} \& {Huchra}(1989)}]{Gell89}
{Geller}, M.~J., \& {Huchra}, J.~P. 1989, Science, 246, 897

\bibitem[{{Giavalisco} {et~al.}(2004){Giavalisco}, {Ferguson}, {Koekemoer},
  {Dickinson}, {Alexander}, {Bauer}, {Bergeron}, {Biagetti}, {Brandt},
  {Casertano}, {Cesarsky}, {Chatzichristou}, {Conselice}, {Cristiani}, {Da
  Costa}, {Dahlen}, {de Mello}, {Eisenhardt}, {Erben}, {Fall}, {Fassnacht},
  {Fosbury}, {Fruchter}, {Gardner}, {Grogin}, {Hook}, {Hornschemeier}, {Idzi},
  {Jogee}, {Kretchmer}, {Laidler}, {Lee}, {Livio}, {Lucas}, {Madau},
  {Mobasher}, {Moustakas}, {Nonino}, {Padovani}, {Papovich}, {Park},
  {Ravindranath}, {Renzini}, {Richardson}, {Riess}, {Rosati}, {Schirmer},
  {Schreier}, {Somerville}, {Spinrad}, {Stern}, {Stiavelli}, {Strolger},
  {Urry}, {Vandame}, {Williams}, \& {Wolf}}]{Giav04}
{Giavalisco}, M., {et~al.} 2004, \apjl, 600, L93

\bibitem[{{Gott} {et~al.}(2005){Gott}, {Juri{\'c}}, {Schlegel}, {Hoyle},
  {Vogeley}, {Tegmark}, {Bahcall}, \& {Brinkmann}}]{Gott05}
{Gott}, J.~R.~I., {Juri{\'c}}, M., {Schlegel}, D., {Hoyle}, F., {Vogeley}, M.,
  {Tegmark}, M., {Bahcall}, N., \& {Brinkmann}, J. 2005, \apj, 624, 463

\bibitem[{{Hall} {et~al.}(2004){Hall}, {Luppino}, {Hodapp}, {Garnett}, {Loose},
  \& {Zandian}}]{Hall04}
{Hall}, D.~N.~B., {Luppino}, G., {Hodapp}, K.~W., {Garnett}, J.~D., {Loose},
  M., \& {Zandian}, M. 2004, in Society of Photo-Optical Instrumentation
  Engineers (SPIE) Conference Series, Vol. 5499, Society of Photo-Optical
  Instrumentation Engineers (SPIE) Conference Series, ed. J.~D. {Garnett} \&
  J.~W. {Beletic}, 1--14

\bibitem[{{Hill} {et~al.}(2010){Hill}, {Driver}, {Cameron}, {Cross}, {Liske},
  \& {Robotham}}]{Hill10}
{Hill}, D.~T., {Driver}, S.~P., {Cameron}, E., {Cross}, N., {Liske}, J., \&
  {Robotham}, A. 2010, \mnras, 404, 1215

\bibitem[{{Huang} {et~al.}(1997){Huang}, {Cowie}, {Gardner}, {Hu}, {Songaila},
  \& {Wainscoat}}]{Huan97}
{Huang}, J.-S., {Cowie}, L.~L., {Gardner}, J.~P., {Hu}, E.~M., {Songaila}, A.,
  \& {Wainscoat}, R.~J. 1997, \apj, 476, 12

\bibitem[{{Huang} {et~al.}(2003){Huang}, {Glazebrook}, {Cowie}, \&
  {Tinney}}]{Huan03}
{Huang}, J.-S., {Glazebrook}, K., {Cowie}, L.~L., \& {Tinney}, C. 2003, \apj,
  584, 203

\bibitem[{{Huchra} {et~al.}(1992){Huchra}, {Geller}, {Clemens}, {Tokarz}, \&
  {Michel}}]{Huch92}
{Huchra}, J.~P., {Geller}, M.~J., {Clemens}, C.~M., {Tokarz}, S.~P., \&
  {Michel}, A. 1992, Bulletin d'Information du Centre de Donnees Stellaires,
  41, 31

\bibitem[{{Inoue} \& {Silk}(2006)}]{Inou06}
{Inoue}, K.~T., \& {Silk}, J. 2006, \apj, 648, 23

\bibitem[{{Jones} {et~al.}(2006){Jones}, {Peterson}, {Colless}, \&
  {Saunders}}]{Jone06}
{Jones}, D.~H., {Peterson}, B.~A., {Colless}, M., \& {Saunders}, W. 2006,
  \mnras, 369, 25

\bibitem[{{Jones} {et~al.}(2004){Jones}, {Saunders}, {Colless}, {Read},
  {Parker}, {Watson}, {Campbell}, {Burkey}, {Mauch}, {Moore}, {Hartley},
  {Cass}, {James}, {Russell}, {Fiegert}, {Dawe}, {Huchra}, {Jarrett}, {Lahav},
  {Lucey}, {Mamon}, {Proust}, {Sadler}, \& {Wakamatsu}}]{Jone04}
{Jones}, D.~H., {et~al.} 2004, \mnras, 355, 747


\bibitem[{{Keenan} {et~al.}(2010{\natexlab{a}}){Keenan}, {Trouille}, {Barger},
  {Cowie}, \& {Wang}}]{Keen10a}
{Keenan}, R.~C., {Trouille}, L., {Barger}, A.~J., {Cowie}, L.~L., \& {Wang},
  W.~H. 2010{\natexlab{a}}, \apjs, 186, 94

\bibitem[{{Keenan} {et~al.}(2010{\natexlab{b}}){Keenan}, {Barger}, {Cowie}, \&
  {Wang}}]{Keen10b}
{Keenan}, R.~C., {Barger}, A.~J., {Cowie}, L.~L., \& {Wang}, W.
  2010{\natexlab{b}}, \apj, 723, 40

\bibitem[{{Kirby} {et~al.}(2008){Kirby}, {Jerjen}, {Ryder}, \&
  {Driver}}]{Kirb08}
{Kirby}, E.~M., {Jerjen}, H., {Ryder}, S.~D., \& {Driver}, S.~P. 2008, \aj,
  136, 1866

\bibitem[{{Kitzbichler} \& {White}(2007)}]{Kitz07}
{Kitzbichler}, M.~G., \& {White}, S.~D.~M. 2007, \mnras, 376, 2

\bibitem[{{Knobel} {et~al.}(2009){Knobel}, {Lilly}, {Iovino}, {Porciani},
  {Kova{\v c}}, {Cucciati}, {Finoguenov}, {Kitzbichler}, {Carollo}, {Contini},
  {Kneib}, {Le F{\`e}vre}, {Mainieri}, {Renzini}, {Scodeggio}, {Zamorani},
  {Bardelli}, {Bolzonella}, {Bongiorno}, {Caputi}, {Coppa}, {de la Torre}, {de
  Ravel}, {Franzetti}, {Garilli}, {Kampczyk}, {Lamareille}, {Le Borgne}, {Le
  Brun}, {Maier}, {Mignoli}, {Pello}, {Peng}, {Perez Montero}, {Ricciardelli},
  {Silverman}, {Tanaka}, {Tasca}, {Tresse}, {Vergani}, {Zucca}, {Abbas},
  {Bottini}, {Cappi}, {Cassata}, {Cimatti}, {Fumana}, {Guzzo}, {Koekemoer},
  {Leauthaud}, {Maccagni}, {Marinoni}, {McCracken}, {Memeo}, {Meneux}, {Oesch},
  {Pozzetti}, \& {Scaramella}}]{Knob09}
{Knobel}, C., {et~al.} 2009, \apj, 697, 1842

\bibitem[{{Kochanek} {et~al.}(2001){Kochanek}, {Pahre}, {Falco}, {Huchra},
  {Mader}, {Jarrett}, {Chester}, {Cutri}, \& {Schneider}}]{Koch01}
{Kochanek}, C.~S., {et~al.} 2001, \apj, 560, 566

\bibitem[{{Kotulla} {et~al.}(2009){Kotulla}, {Fritze}, {Weilbacher}, \&
  {Anders}}]{Kotu09}
{Kotulla}, R., {Fritze}, U., {Weilbacher}, P., \& {Anders}, P. 2009, \mnras,
  396, 462

\bibitem[{{Kova{\v c}} {et~al.}(2010){Kova{\v c}}, {Lilly}, {Cucciati},
  {Porciani}, {Iovino}, {Zamorani}, {Oesch}, {Bolzonella}, {Knobel},
  {Finoguenov}, {Peng}, {Carollo}, {Pozzetti}, {Caputi}, {Silverman}, {Tasca},
  {Scodeggio}, {Vergani}, {Scoville}, {Capak}, {Contini}, {Kneib}, {Le
  F{\`e}vre}, {Mainieri}, {Renzini}, {Bardelli}, {Bongiorno}, {Coppa}, {de la
  Torre}, {de Ravel}, {Franzetti}, {Garilli}, {Guzzo}, {Kampczyk},
  {Lamareille}, {Le Borgne}, {Le Brun}, {Maier}, {Mignoli}, {Pello}, {Perez
  Montero}, {Ricciardelli}, {Tanaka}, {Tresse}, {Zucca}, {Abbas}, {Bottini},
  {Cappi}, {Cassata}, {Cimatti}, {Fumana}, {Koekemoer}, {Maccagni}, {Marinoni},
  {McCracken}, {Memeo}, {Meneux}, \& {Scaramella}}]{Kova10}
{Kova{\v c}}, K., {et~al.} 2010, \apj, 708, 505

\bibitem[{{Kron}(1980)}]{Kron80}
{Kron}, R.~G. 1980, \apjs, 43, 305

\bibitem[{{Kurtz} \& {Mink}(1998)}]{Kurt98}
{Kurtz}, M.~J., \& {Mink}, D.~J. 1998, \pasp, 110, 934

\bibitem[{{Lawrence} {et~al.}(2007){Lawrence}, {Warren}, {Almaini}, {Edge},
  {Hambly}, {Jameson}, {Lucas}, {Casali}, {Adamson}, {Dye}, {Emerson},
  {Foucaud}, {Hewett}, {Hirst}, {Hodgkin}, {Irwin}, {Lodieu}, {McMahon},
  {Simpson}, {Smail}, {Mortlock}, \& {Folger}}]{Lawr07}
{Lawrence}, A., {et~al.} 2007, \mnras, 379, 1599

\bibitem[{{Le F{\`e}vre} {et~al.}(2005){Le F{\`e}vre}, {Vettolani}, {Garilli},
  {Tresse}, {Bottini}, {Le Brun}, {Maccagni}, {Picat}, {Scaramella},
  {Scodeggio}, {Zanichelli}, {Adami}, {Arnaboldi}, {Arnouts}, {Bardelli},
  {Bolzonella}, {Cappi}, {Charlot}, {Ciliegi}, {Contini}, {Foucaud},
  {Franzetti}, {Gavignaud}, {Guzzo}, {Ilbert}, {Iovino}, {McCracken}, {Marano},
  {Marinoni}, {Mathez}, {Mazure}, {Meneux}, {Merighi}, {Paltani}, {Pell{\`o}},
  {Pollo}, {Pozzetti}, {Radovich}, {Zamorani}, {Zucca}, {Bondi}, {Bongiorno},
  {Busarello}, {Lamareille}, {Mellier}, {Merluzzi}, {Ripepi}, \&
  {Rizzo}}]{Lefe05}
{Le F{\`e}vre}, O., {et~al.} 2005, \aap, 439, 845

\bibitem[{{Lilly} {et~al.}(1991){Lilly}, {Cowie}, \& {Gardner}}]{Lill91}
{Lilly}, S.~J., {Cowie}, L.~L., \& {Gardner}, J.~P. 1991, \apj, 369, 79

\bibitem[{{Lilly} {et~al.}(2009){Lilly}, {Le Brun}, {Maier}, {Mainieri},
  {Mignoli}, {Scodeggio}, {Zamorani}, {Carollo}, {Contini}, {Kneib}, {Le
  F{\`e}vre}, {Renzini}, {Bardelli}, {Bolzonella}, {Bongiorno}, {Caputi},
  {Coppa}, {Cucciati}, {de la Torre}, {de Ravel}, {Franzetti}, {Garilli},
  {Iovino}, {Kampczyk}, {Kovac}, {Knobel}, {Lamareille}, {Le Borgne}, {Pello},
  {Peng}, {P{\'e}rez-Montero}, {Ricciardelli}, {Silverman}, {Tanaka}, {Tasca},
  {Tresse}, {Vergani}, {Zucca}, {Ilbert}, {Salvato}, {Oesch}, {Abbas},
  {Bottini}, {Capak}, {Cappi}, {Cassata}, {Cimatti}, {Elvis}, {Fumana},
  {Guzzo}, {Hasinger}, {Koekemoer}, {Leauthaud}, {Maccagni}, {Marinoni},
  {McCracken}, {Memeo}, {Meneux}, {Porciani}, {Pozzetti}, {Sanders},
  {Scaramella}, {Scarlata}, {Scoville}, {Shopbell}, \& {Taniguchi}}]{Lill09}
{Lilly}, S.~J., {et~al.} 2009, \apjs, 184, 218

\bibitem[{{Liske} {et~al.}(2003){Liske}, {Lemon}, {Driver}, {Cross}, \&
  {Couch}}]{Lisk03}
{Liske}, J., {Lemon}, D.~J., {Driver}, S.~P., {Cross}, N.~J.~G., \& {Couch},
  W.~J. 2003, \mnras, 344, 307

\bibitem[{{Lockman} {et~al.}(1986){Lockman}, {Jahoda}, \& {McCammon}}]{Lock86}
{Lockman}, F.~J., {Jahoda}, K., \& {McCammon}, D. 1986, \apj, 302, 432

\bibitem[{{Lonsdale} {et~al.}(2003){Lonsdale}, {Smith}, {Rowan-Robinson},
  {Surace}, {Shupe}, {Xu}, {Oliver}, {Padgett}, {Fang}, {Conrow},
  {Franceschini}, {Gautier}, {Griffin}, {Hacking}, {Masci}, {Morrison},
  {O'Linger}, {Owen}, {P{\'e}rez-Fournon}, {Pierre}, {Puetter}, {Stacey},
  {Castro}, {Polletta}, {Farrah}, {Jarrett}, {Frayer}, {Siana}, {Babbedge},
  {Dye}, {Fox}, {Gonzalez-Solares}, {Salaman}, {Berta}, {Condon}, {Dole}, \&
  {Serjeant}}]{Lons03}
{Lonsdale}, C.~J., {et~al.} 2003, \pasp, 115, 897

\bibitem[{{Lynden-Bell}(1971)}]{Lynd71}
{Lynden-Bell}, D. 1971, \mnras, 155, 95

\bibitem[{{Mannucci} {et~al.}(2001){Mannucci}, {Basile}, {Poggianti},
  {Cimatti}, {Daddi}, {Pozzetti}, \& {Vanzi}}]{Mann01}
{Mannucci}, F., {Basile}, F., {Poggianti}, B.~M., {Cimatti}, A., {Daddi}, E.,
  {Pozzetti}, L., \& {Vanzi}, L. 2001, \mnras, 326, 745

\bibitem[{{Maraston} {et~al.}(2006){Maraston}, {Daddi}, {Renzini}, {Cimatti},
  {Dickinson}, {Papovich}, {Pasquali}, \& {Pirzkal}}]{Mara06}
{Maraston}, C., {Daddi}, E., {Renzini}, A., {Cimatti}, A., {Dickinson}, M.,
  {Papovich}, C., {Pasquali}, A., \& {Pirzkal}, N. 2006, \apj, 652, 85

\bibitem[{{Markwardt}(2009)}]{Mark09}
{Markwardt}, C.~B. 2009, in Astronomical Society of the Pacific Conference
  Series, Vol. 411, Astronomical Data Analysis Software and Systems XVIII, ed.
  {D.~A.~Bohlender, D.~Durand, \& P.~Dowler}, 251

\bibitem[{{Marra} \& {Notari}(2011)}]{Marr11}
{Marra}, V., \& {Notari}, A. 2011, Classical and Quantum Gravity, 28, 164004

\bibitem[{{Marra} \& {P{\"a}{\"a}kk{\"o}nen}(2010)}]{Marr10}
{Marra}, V., \& {P{\"a}{\"a}kk{\"o}nen}, M. 2010, JCAP, 12, 21

\bibitem[{{McCracken} {et~al.}(2007){McCracken}, {Peacock}, {Guzzo}, {Capak},
  {Porciani}, {Scoville}, {Aussel}, {Finoguenov}, {James}, {Kitzbichler},
  {Koekemoer}, {Leauthaud}, {Le F{\`e}vre}, {Massey}, {Mellier}, {Mobasher},
  {Norberg}, {Rhodes}, {Sanders}, {Sasaki}, {Taniguchi}, {Thompson}, {White},
  \& {El-Zant}}]{Mccr07}
{McCracken}, H.~J., {et~al.} 2007, \apjs, 172, 314

\bibitem[{{Moss} {et~al.}(2011){Moss}, {Zibin}, \& {Scott}}]{Moss11}
{Moss}, A., {Zibin}, J.~P., \& {Scott}, D. 2011, \prd, 83, 103515

\bibitem[{{Page} \& {Carrera}(2000)}]{Page2000}
{Page}, M.~J., \& {Carrera}, F.~J. 2000, \mnras, 311, 433

\bibitem[{{Riess} {et~al.}(2011){Riess}, {Macri}, {Casertano}, {Lampeitl},
  {Ferguson}, {Filippenko}, {Jha}, {Li}, \& {Chornock}}]{Ries11}
{Riess}, A.~G., {et~al.} 2011, \apj, 730, 119

\bibitem[{{Sandage} {et~al.}(1979){Sandage}, {Tammann}, \& {Yahil}}]{Sand79}
{Sandage}, A., {Tammann}, G.~A., \& {Yahil}, A. 1979, \apj, 232, 352

\bibitem[{{Schechter}(1976)}]{Sche76}
{Schechter}, P. 1976, \apj, 203, 297

\bibitem[{{Schmidt}(1968)}]{Schm68}
{Schmidt}, M. 1968, \apj, 151, 393

\bibitem[{{Scoville} {et~al.}(2007{\natexlab{a}}){Scoville}, {Aussel},
  {Benson}, {Blain}, {Calzetti}, {Capak}, {Ellis}, {El-Zant}, {Finoguenov},
  {Giavalisco}, {Guzzo}, {Hasinger}, {Koda}, {Le F{\`e}vre}, {Massey},
  {McCracken}, {Mobasher}, {Renzini}, {Rhodes}, {Salvato}, {Sanders}, {Sasaki},
  {Schinnerer}, {Sheth}, {Shopbell}, {Taniguchi}, {Taylor}, \&
  {Thompson}}]{Scov07b}
{Scoville}, N., {et~al.} 2007{\natexlab{a}}, \apjs, 172, 150

\bibitem[{{Scoville} {et~al.}(2007{\natexlab{b}}){Scoville}, {Aussel}, {Brusa},
  {Capak}, {Carollo}, {Elvis}, {Giavalisco}, {Guzzo}, {Hasinger}, {Impey},
  {Kneib}, {LeFevre}, {Lilly}, {Mobasher}, {Renzini}, {Rich}, {Sanders},
  {Schinnerer}, {Schminovich}, {Shopbell}, {Taniguchi}, \& {Tyson}}]{Scov07a}
---. 2007{\natexlab{b}}, \apjs, 172, 1

\bibitem[{{Skrutskie} {et~al.}(2006){Skrutskie}, {Cutri}, {Stiening},
  {Weinberg}, {Schneider}, {Carpenter}, {Beichman}, {Capps}, {Chester},
  {Elias}, {Huchra}, {Liebert}, {Lonsdale}, {Monet}, {Price}, {Seitzer},
  {Jarrett}, {Kirkpatrick}, {Gizis}, {Howard}, {Evans}, {Fowler}, {Fullmer},
  {Hurt}, {Light}, {Kopan}, {Marsh}, {McCallon}, {Tam}, {Van Dyk}, \&
  {Wheelock}}]{Skru06}
{Skrutskie}, M.~F., {et~al.} 2006, \aj, 131, 1163

\bibitem[{{Smith} {et~al.}(2009){Smith}, {Loveday}, \& {Cross}}]{Smit09}
{Smith}, A.~J., {Loveday}, J., \& {Cross}, N.~J.~G. 2009, \mnras, 397, 868

\bibitem[{{Springel} {et~al.}(2005){Springel}, {White}, {Jenkins}, {Frenk},
  {Yoshida}, {Gao}, {Navarro}, {Thacker}, {Croton}, {Helly}, {Peacock}, {Cole},
  {Thomas}, {Couchman}, {Evrard}, {Colberg}, \& {Pearce}}]{Spri05}
{Springel}, V., {et~al.} 2005, \nat, 435, 629

\bibitem[{{Steffen} {et~al.}(2004){Steffen}, {Barger}, {Capak}, {Cowie},
  {Mushotzky}, \& {Yang}}]{Stef04}
{Steffen}, A.~T., {Barger}, A.~J., {Capak}, P., {Cowie}, L.~L., {Mushotzky},
  R.~F., \& {Yang}, Y. 2004, \aj, 128, 1483

\bibitem[{{Tonry} \& {Davis}(1979)}]{Tonr79}
{Tonry}, J., \& {Davis}, M. 1979, \aj, 84, 1511

\bibitem[{{Trouille} {et~al.}(2008){Trouille}, {Barger}, {Cowie}, {Yang}, \&
  {Mushotzky}}]{Trou08}
{Trouille}, L., {Barger}, A.~J., {Cowie}, L.~L., {Yang}, Y., \& {Mushotzky},
  R.~F. 2008, \apjs, 179, 1

\bibitem[{{Trouille} {et~al.}(2009){Trouille}, {Barger}, {Cowie}, {Yang}, \&
  {Mushotzky}}]{Trou09}
---. 2009, \apj, 703, 2160

\bibitem[{{Wang} {et~al.}(2010){Wang}, {Cowie}, {Barger}, {Keenan}, \&
  {Ting}}]{Wang10}
{Wang}, W., {Cowie}, L.~L., {Barger}, A.~J., {Keenan}, R.~C., \& {Ting}, H.
  2010, \apjs, 187, 251

\bibitem[{{Willmer}(1997)}]{Will97}
{Willmer}, C.~N.~A. 1997, \aj, 114, 898

\bibitem[{{Yang} {et~al.}(2004){Yang}, {Mushotzky}, {Steffen}, {Barger}, \&
  {Cowie}}]{Yang04}
{Yang}, Y., {Mushotzky}, R.~F., {Steffen}, A.~T., {Barger}, A.~J., \& {Cowie},
  L.~L. 2004, \aj, 128, 1501

\bibitem[{{York} {et~al.}(2000){York}, {Adelman}, {Anderson}, {Anderson},
  {Annis}, {Bahcall}, {Bakken}, {Barkhouser}, {Bastian}, {Berman}, {Boroski},
  {Bracker}, {Briegel}, {Briggs}, {Brinkmann}, {Brunner}, {Burles}, {Carey},
  {Carr}, {Castander}, {Chen}, {Colestock}, {Connolly}, {Crocker}, {Csabai},
  {Czarapata}, {Davis}, {Doi}, {Dombeck}, {Eisenstein}, {Ellman}, {Elms},
  {Evans}, {Fan}, {Federwitz}, {Fiscelli}, {Friedman}, {Frieman}, {Fukugita},
  {Gillespie}, {Gunn}, {Gurbani}, {de Haas}, {Haldeman}, {Harris}, {Hayes},
  {Heckman}, {Hennessy}, {Hindsley}, {Holm}, {Holmgren}, {Huang}, {Hull},
  {Husby}, {Ichikawa}, {Ichikawa}, {Ivezi{\'c}}, {Kent}, {Kim}, {Kinney},
  {Klaene}, {Kleinman}, {Kleinman}, {Knapp}, {Korienek}, {Kron}, {Kunszt},
  {Lamb}, {Lee}, {Leger}, {Limmongkol}, {Lindenmeyer}, {Long}, {Loomis},
  {Loveday}, {Lucinio}, {Lupton}, {MacKinnon}, {Mannery}, {Mantsch}, {Margon},
  {McGehee}, {McKay}, {Meiksin}, {Merelli}, {Monet}, {Munn}, {Narayanan},
  {Nash}, {Neilsen}, {Neswold}, {Newberg}, {Nichol}, {Nicinski}, {Nonino},
  {Okada}, {Okamura}, {Ostriker}, {Owen}, {Pauls}, {Peoples}, {Peterson},
  {Petravick}, {Pier}, {Pope}, {Pordes}, {Prosapio}, {Rechenmacher}, {Quinn},
  {Richards}, {Richmond}, {Rivetta}, {Rockosi}, {Ruthmansdorfer}, {Sandford},
  {Schlegel}, {Schneider}, {Sekiguchi}, {Sergey}, {Shimasaku}, {Siegmund},
  {Smee}, {Smith}, {Snedden}, {Stone}, {Stoughton}, {Strauss}, {Stubbs},
  {SubbaRao}, {Szalay}, {Szapudi}, {Szokoly}, {Thakar}, {Tremonti}, {Tucker},
  {Uomoto}, {Vanden Berk}, {Vogeley}, {Waddell}, {Wang}, {Watanabe},
  {Weinberg}, {Yanny}, \& {Yasuda}}]{York00}
{York}, D.~G., {et~al.} 2000, \aj, 120, 1579

\bibitem[{{Zhang} \& {Stebbins}(2011)}]{Zhan11}
{Zhang}, P., \& {Stebbins}, A. 2011, Physical Review Letters, 107, 041301

\bibitem[{{Zibetti} {et~al.}(2012){Zibetti}, {Gallazzi}, {Charlot}, {Pasquali},
  \& {Pierini}}]{Zibb12}
{Zibetti}, S., {Gallazzi}, A., {Charlot}, S., {Pasquali}, A., \& {Pierini}, D.
  2012, ArXiv e-prints

\bibitem[{{Zibin} {et~al.}(2008){Zibin}, {Moss}, \& {Scott}}]{Zibi08}
{Zibin}, J.~P., {Moss}, A., \& {Scott}, D. 2008, Physical Review Letters, 101,
  251303

\end{thebibliography}
